\documentclass[journal,12pt]{IEEEtran}

\usepackage{cite}

\ifCLASSINFOpdf
\else
\fi
\usepackage{url}

\usepackage{tabularx}
\usepackage{xcolor}
\usepackage[hidelinks,colorlinks,unicode]{hyperref}
\usepackage{graphicx,wrapfig,amssymb,amsfonts,amsmath,epstopdf,array}
\usepackage{array}
\usepackage{subcaption}
\usepackage{cleveref}
\usepackage{ulem}
\usepackage{longtable}
\usepackage{algorithmic}
\usepackage{algorithm}
\usepackage{multirow}
\usepackage{multicol}
\usepackage{comment}
\usepackage{soul}
\newcommand\numberthis{\addtocounter{equation}{1}\tag{\theequation}}
\newcolumntype{P}[1]{>{\centering\arraybackslash}p{#1}}

\begin{document}
%
\title{Indoor Position and Attitude Tracking with SO(3) Manifold}
%
%
%

\author{{Hammam~Salem,
        Mohanad~Ahmed, Mohammed AlSharif,
        Ali~Muqaibel,~\IEEEmembership{Senior~Member,~IEEE}
        and~Tareq~Al-Naffouri,~\IEEEmembership{Senior~Member,~IEEE}}
 \thanks{Hammam Salem was with Center for Communication Systems and Sensing,
Electrical Engineering Department, King Fahd University of Petroleum and Minerals, Dhahran 31261, KSA (e-mail: hmmam1413@gmail.com).}
 \thanks{Mohanad Ahmed was with Information Science Lab, King Abdullah University of Science and Technology, Thuwal 23955-6900, KSA (e-mail: m.a.m.elhassan@gmail.com).}
\thanks{Mohammed AlSharif is with aramco research center at kaust, Thuwal 23955-6900, KSA (e-mail: mohammed.alsharif@kaust.edu.sa).}
\thanks{Ali Muqaibel is with Center for Communication Systems and Sensing,
Electrical Engineering Department, King Fahd University of Petroleum and Minerals, Dhahran 31261, KSA (e-mail: muqaibel@kfupm.edu.sa).}
\thanks{Tareq Al-Naffouri is with Information Science Lab, King Abdullah University of Science and Technology, Thuwal 23955-6900, KSA (e-mail: tareq.alnaffouri@kaust.edu.sa).}}
\maketitle

\begin{abstract}

Driven by technological breakthroughs, indoor tracking and localization have gained importance in various applications including the Internet of Things (IoT), robotics, and unmanned aerial vehicles (UAVs). 
To tackle some of the challenges associated with indoor tracking, this study explores the potential benefits of incorporating the SO($3$) manifold structure of the rotation matrix. The goal is to enhance the 3D tracking performance of the extended Kalman filter (EKF) and unscented Kalman filter (UKF) of a moving target within an indoor environment. 
Our results demonstrate that the proposed extended Kalman filter with Riemannian (EKFRie) and unscented Kalman filter with Riemannian (UKFRie) algorithms consistently outperform the conventional EKF and UKF in terms of position and orientation accuracy. While the conventional EKF and UKF achieved root mean square error (RMSE) of $0.36$m and $0.43$m, respectively, for a long stair path, the proposed EKFRie and UKFRie algorithms achieved a lower RMSE of $0.21$m and $0.10$m. 
Our results show also the outperforming of the proposed algorithms over the EKF and UKF algorithms with the Isosceles triangle manifold. While the latter achieved RMSE of $ 7.26$cm and $7.27$cm, respectively, our proposed algorithms achieved RMSE of $6.73$cm and $6.16$cm. These results demonstrate the enhanced performance of the proposed algorithms. 

\end{abstract}


%
\IEEEpeerreviewmaketitle

\section{Introduction}
%
%
%
%
\newcommand{\question}[1]{
}
\newcommand{\answer}[1]{\textit{#1}}
\newcommand{\editorial}[1]{
}
\newcommand{\practice}[1]{
}
\newcommand{\hammam}[1]{\footnote{\textcolor{green}{#1}}}
\newcommand{\Sout}[1]{
}
\newcommand{\sub}[1]{\textcolor{purple}{#1}}


With the technological progress that the world is witnessing, positioning and tracking appear as important factors \cite{khudhair2016wireless}. Positioning (or localization) means estimating the coordinates and orientation of an object when it is stationary. If the object is moving, the process of estimating its changing position and orientation is known as tracking. However, if the context provides clarity, it is sometimes acceptable to use positioning and tracking interchangeably.
 
Many applications of the current technological progress rely on positioning and tracking. 
The Internet of Things (IoT) is one of the prominent areas of research \cite{8073629} where the rapid development has created a heightened demand for efficient positioning and tracking systems \cite{lin2016human}. Smart parking \cite{khanna2016iot}, smart cities \cite{alsinglawi2017rfid}, and asset Management \cite{lee2019bluetooth} are some examples of localization-based IoT applications. 

The robotics revolution is another example where an enormous number of researchers are working to boost the robot's role in human life. But to effectively control these smart assistants, we need to accurately track their positions and orientations \cite{1041428}. 
The third example that shows the importance of positioning and tracking is Unmanned Aerial Vehicles (UAVs) and drone activities due to their roles in national security, surveillance, media, entertainment, etc. Enhancement in positioning and tracking systems will encourage additional applications. Even in delicate surgical operations, robot positioning plays a vital role in ensuring accuracy and safety \cite{peng2020real}. As robots continue to evolve, precise localization will be the cornerstone of their success.

Various technologies can be utilized for localization, including Bluetooth, WiFi, ultrasound, and optics. Multiple techniques are available for range estimation based on transmitted signals, such as time of arrival (ToA), time difference of arrival (TDoA), angle of arrival (AoA), and received signal strength (RSS). For this work, we assume that the estimated ranges are readily accessible, and therefore, we will not delve into these specific technologies and techniques. For more detailed information on these technologies and techniques, see \cite{zafari2019survey}.

This work deals with the indoor scenario where the tracking depends mainly on systems that serve limited areas like ultrasonic systems (USS) and built-in sensors like inertial measurement units (IMU). This kind of tracking is challenging and it is a trend research topic with no prevailing solution so far \cite{AB7042271}. Some of the indoor tracking challenges include the lack of reliable Global Positioning System (GPS) signals due to signal penetration issues \cite{alhafnawi2023survey}, multipath caused by walls and furniture \cite{gutmann2013challenges}, and the demand for high accuracy, which is often in decimeter-level \cite{zafari2019survey}.

It is possible to employ either a single or multiple transceivers to locate the object. In literature, many works have used multiple receivers for various purposes, including reduced computation complexity \cite{busanelli2010uwb, correa2016indoor}, simplified synchronization \cite{park2021single, shalaby2022ultra}, and improved tracking accuracy \cite{namazi2022improving}. Alejandro Correa \textit{et al.} \cite{correa2016indoor} used a multiple-receivers approach combined with machine learning (ML) algorithms to track a  pedestrian in the indoor environment. Their work achieved $90\%$ of root mean square error (RMSE) below $2$m and $2.8$m for areas around $100m^2$ and $533.12m^2$ respectively.  
Multiple optical receivers were used also by \cite{yasir2015indoor} for tracking in the indoor scenario and achieved position error around $0.17$m at large luminous flux. The root mean square error (RMSE) of $0.89$m is achieved by utilizing the Kalman filter (KF) and machine learning \cite{mahfouz2014target}.

The tracking performance is influenced by a number of factors, including the formulation of the optimization problem and its constraints. Since the rotation matrix belongs to the SO($n$) manifold, a subgroup of the Riemannian manifold, this work opts to express angles of rotation in the form of rotation matrix and track the matrix itself. The purpose is to investigate how adopting SO(3) manifold structure can aid in improving indoor tracking accuracy when the isosceles triangular transmitter configuration is used in the setting. Three transmitters are employed for both orientation estimation and enhanced position accuracy. This work leverages recent advances in Riemannian geometry \cite{Manopt, boumal2020introduction, AbseOptimizationAlgorithmsl, 7528889} to apply Riemannian optimization in the tracking problem. The extended Kalman filter (EKF) and unscented Kalman filter (UKF) are used for comparison purposes. The contributions of this work can be outlined as follows:

\begin{enumerate}
    \item Augmenting the conventional EKF and UKF with Riemannian tools, by means of SO(3) manifold, to enhance the performance (accuracy). Our results show that the accuracy is indeed improved,
      \item Evaluating the proposed algorithms over different measurement variances and IMU data rates,
   \item Showing that the gain in performance is negligible if the manifold tools, by means of isosceles triangle manifold, are applied only to the measurements. 
   \end{enumerate}

The rest of this paper is organized as follows: Section \ref{sec:LitReview} provides the reader with a literature review about enhancing the performance of KFs, localization by using Riemannian geometry, and merging KF with Riemannian optimization to handle tracking problems. 
The essential Riemannian tools, system model and problem formulation are presented in detail in Section \ref{sec:sys.Model}. Section \ref{sec:proposed.Algor} discusses the proposed algorithms: EKF with Riemannian optimization (EKFRie) and UKF with Riemannian optimization (UKFRie).
Finally, Sections \ref{sec:ResulandDiss} and \ref{sec:Concl} discuss the results and present the conclusions of this work. This work is partially based on some content of the thesis \cite{Salem2024position}.

\section{Literature Review}
\label{sec:LitReview}
Moving target tracking problems appeared in the middle of the twentieth century to enable radars to track missiles \cite{delano1953theory, BallTrackerTpeeds_}. The work continued to evolve until 1960 when R. E. Kalman published his paper \cite{kalman1960new} which provided a very vital linear estimator, known as Kalman filter (KF),  to estimate the linearly modeled hidden states. Since then, KF has become a widely used estimator in tracking. EKF and UKF \cite{julier1997new, van2004sigma, wan2000unscented, bar2001estimation, sayed_2022} appeared as nonlinear versions of KF to deal with nonlinear systems. Although KF and its nonlinear versions are widely used in tracking problems because of their high efficiency and accuracy in linear and weakly nonlinear models, they are strongly affected by the behavior of the moving target and the nonlinearity degree of the model.

This literature review examines recent attempts to enhance indoor positioning performance. It focuses on three main approaches: modifications to the Kalman filter, the use of Riemannian optimization, and the fusion of Kalman filters with Riemannian manifolds. 

In the literature, many techniques have been used to enhance KF's performance in tracking. For instance, the authors in \cite{liu2014modified} tackled the problem of high maneuvering by modifying Kalman gain and scaling the state noise covariance by an estimated factor. On the other hand, in an attempt to address the deterioration in EKF performance caused by increased system nonlinearity and model uncertainty, \cite{wang2018variational} used a fifth-degree interpolatory Cubature Kalman Filter (ICKF) to get acceptable performance at the abrupt change of states. 
 
 In another direction, \cite{agarwal2022neural, revach2022kalmannet, millidge2021neural, mahfouz2014target} used machine learning to improve the tracking performance of different forms of Kalman filter. For instance, \cite{mahfouz2014target} used radio fingerprints of received signal strength indicators as inputs to the machine learning (ML) algorithm to estimate the target position. Then a KF uses these estimates to update the estimated position based on acceleration and velocity. Some works used an analytical approach to overcome the sensitivity and complexity of the inverse operation of the covariance matrix in the Kalman filter. For instance, \cite{zhou2019new} extracted singular value decomposition (SVD) of the state covariance and used it rather than the original state covariance to calculate the Kalman gain to overcome the matrix inversion sensitivity. To avoid the violation of the positive definite condition of the covariance matrix, the authors replaced the covariance matrix with its $\mathbf{QR}$ decomposition where $\mathbf{Q}$ is an orthonormal matrix and $\mathbf{R}$ is an upper triangular matrix. 

 On the other hand, there has been continuous development in the field of differential geometry, particularly in the study of manifolds, since the introduction of Riemannian manifolds by Bernhard Riemann. Some researchers have leveraged the Riemannian structure in tracking applications to enhance performance \cite{bicanic2021multi, yun2013multi, wu2008probabilistic, mashtakov2017tracking, khan2011bayesian, khan2011tracking}. Riemannian geometry presents challenges due to its limited tools compared to Euclidean space. However, Ref. \cite{bicanic2021multi} extended the tracking problems from Euclidean space to Riemannian manifold space to improve the performance by utilizing probabilistic data association and leveraging Riemannian tools. 
 In a similar vein, Ref. \cite{yun2013multi} applied a manifold-based maximum likelihood (ML) estimator to estimate the target location while considering geometric constraints.

\cite[Hertzberg]{hertzberg2013integrating} and \cite[Hauberg]{hauberg2013unscented} discussed applying UKF with Riemannian manifold. Hertzberg focused on generalizing the needed mathematical operations in sensor fusion algorithms, such as summation and subtraction, to make them applicable in Riemannian space. Rather than that, Hauberg generalized the unscented transformation to be effective in both Riemannian and Euclidean environments. In this context, \cite{menegaz2018unscented} also provided Riemannian-based UKF by extending the sigma point from Euclidean space into the Riemannian manifold in a closed form.

 As our work introduces a target equipped with three transmitters, this review also includes a comparison with literature related to localization and tracking using multiple transmitters/receivers. Elnaz Namaz \textit{et al.} \cite{namazi2022improving} focused on estimating the target location by merging two GPS receivers with map matching. However, this work limited its investigation to the intersection and straight road scenarios, which restricts the generalizability of their findings.
 Ref. \cite{correa2016indoor} employed EKF to merge signal strength-based readings from wireless sensor networks and showed the advantage of multiple receiver over one receiver in terms of accuracy. 
 Ultra-wideband signal (UWB) and IMU were used in \cite{cao2021itracku} to track pen-like instruments. UWB signal was used to localize and transfer IMU's information simultaneously which saves resources. The UWB and IMU-based system achieved  $90\%$ error within $7$mm for a $5m^2$ area. However, as the work was evaluated in small areas, its applicability and validity for larger areas may be the subject of further investigation. Ref. \cite{krishnaveni2022indoor} combined readings from an IMU and UWB technology using EKF and UKF for indoor tracking. However, their algorithms exhibited position error rates of $75\%$ and $90\%$ within $2$m, respectively, which may not be suitable for many applications. Also, two UWB receivers and an odometer system were adopted by Ref. \cite{busanelli2010uwb} to estimate the position and orientation of a moving target by applying EKF. Two receivers were used to determine the heading angle with RMSE of less than $0.8$m.

 Focusing on acoustic-based indoor localization, Ruizhi Chen \textit{et al.} \cite{chen2021precise} introduced a least squares (LS) algorithm based on time-difference-of-arrival (TDoA) for static localization and the EKF algorithm with IMU for dynamic tracking. However, their outcomes showed relatively low accuracy for many indoor applications, with $95\%$ of the error falling within $1.6 $m. 
 Al-Sharif \textit{et al.} \cite{alsharif2024kalman} enhanced EKF and UKF for indoor tracking using a three-receiver isosceles triangle setup. While \cite{alsharif2024kalman} employed the isosceles triangle manifold, our work primarily utilizes SO(3). Al-Sharif applied retractions once at the end of the measurement update whilst we applied it twice: after the time and measurement update. Moreover, unlike Al-Sharif's approach, we ensure the covariance matrix is parallel transported across the manifold, reflecting the Riemannian steps. We borrowed this idea from \cite{hauberg2013unscented}.

  Table \ref{tab1} presents a comparison of some indoor localization and tracking studies, providing additional details about the algorithms used, noise statistics, evaluation techniques, and any identified limitations. As the table shows, most of the works covered are based on EKF/UKF. 

\begin{table*}[!t]
\centering
    \caption{Comparison between some indoor localization works  featuring multiple receivers}
\begin{tabular}{|m{2em}|m{17 em}|m{11em}|m{10em}|m{11em}|}
\hline
  Ref. & \centering Algorithm & \centering Noise Statistics & \centering Evaluation & \hspace{3 em} Shortcomings \\

  \hline
\cite{lee2019bluetooth} & Trilateration with RSS of Bluetooth signal & Paspberry Pi $3$ model B receiver. Bluetooth: $4.1$ classic, BLE & RMSE = $2.13 $m & Low accuracy for many indoor applications \\
  \hline
\cite{busanelli2010uwb} & EKF with odometer readings and TDoA-based ranges by using UWB system & $\sigma^2_{state}$ = $0.1mm^2$ and $3.06 \mu$ rad$^2$ for position and orientation & Position RMSE $<0.8$m, orientation RMSE $<0.5 $rad &  Requiring precise synchronization \\
\hline
\cite{correa2016indoor} &  EKF with position, heading, and speed from ML algorithms & 
    $\sigma^2_{state}$: position $=5$ , speed = $0.5$

  & 
  RMSE = $1$m and $0.75$m for path = $44$m and $74.4$m  &  
  Assumed constant velocity
 \\
\hline 
\cite{yasir2015indoor} & Received light intensities and accelerometer readings. & $\sigma^2_{accelerometer}= 5.7\times10^{-5} (m/s^2)^2$ & Mean error $\approx0.06$m
& Didn't provide MSE, required LOS. \\
\hline
\cite{cao2021itracku} & KF with IMU readings and TDoA-based ranges by using UWB system& 9DoF IMU and DW1000 UWB chip & Median error $\approx2.9 $mm  
& Examined over a small area ($5m^2$)  \\

\hline
\cite{krishnaveni2022indoor}& EKF \& UKF with IMU readings and ToA-based ranges by using UWB system  & $\sigma^{2}_{meas.}= 1\times10^{-4}m^2$

 $\sigma^{2}_{process}$ =  $1\times10^{-10}m^2$ & $90\%$ and $75\%$ of \scriptsize{UKF} 
 and EKF error within $2$m & Examined over one path shape, low accuracy \\
\hline 
\cite{chen2021precise} & EKF with IMU readings and ToA-based ranges by using UWB system & None mentioned  & $95\%$ of the error bellow $1.6$m & Examined $2D$ only, low accuracy \\
\hline
\cite{qi2017robust} & Least square algorithm based on ToF measurements. & None mentioned &  Maximum error = $10.24 $mm & Considered one trajectory shape with a small distance \\
\hline
\cite{cai2019asynchronous} & Particle filters with acoustic Doppler velocity and beacon timestamps  & \scriptsize{INMP411} microphone and raspberry Pi $3$ board & $90\%$ of the error within $0.49 $m &  Acoustic's low coverage necessitates more hardware \\
\hline
    \end{tabular}
    \label{tab1}
\end{table*}

 Based on the literature review conducted, there is a lack of exploiting the SO($3$) structure of the rotation matrix to enhance the tracking performance of EKF and UKF. Moreover, the impact of exploiting the receivers' isosceles triangle geometry at the measurement stages, before applying Kalman filter, has not been thoroughly discussed in the literature to the best of our knowledge. Therefore, this work aims to address these gabs by utilizing EKF and UKF algorithms with Riemannian optimization.

\section{System Model and Problem Formulation}
\label{sec:sys.Model}
In this section, we describe the considered system and the mathematical formulation of the tracking problem. But to provide necessary context, a brief summary of the needed Riemannian tools is presented first. For a more in-depth exploration of Riemannian geometry, please refer to \cite{boumal2020introduction, AbseOptimizationAlgorithmsl}. 
\subsection{Riemannian Tools}
Since this work relies particularly on the SO($3$) manifold, this section focuses on that manifold and its definition and mathematical operators.
\newtheorem{theorem}{\textbf{Definition}}
        \begin{theorem}
            SO($3$) manifold is a set of square matrices with size ($3\times3$) embedded in $\mathbb{R}^{3 \times 3}$ (where $\mathbb{R}^{3\times3}$ is the embedding space endowed with the inner product) have orthonormal columns and their determinant equal to (+1). 
        \end{theorem}
      
     This definition can be stated equivalently as:
     $ \text{SO}(3)=\{ X \in \mathbb{R}^{3\times3}: X^{\top}X=XX^{\top}=I_3 \text{ and }$ $ det(X)=+1 \} $.
     Table \ref{tab:geometry for SO(3)} shows the formulas of the required geometrical tools for SO($3$). For more about the Riemannian geometry of SO($3$), see \cite{Manopt}.
       \begin{table}[!ht]
          \caption{Riemannian geometry for SO($3$) manifold}

           \centering
           \begin{tabular}{ |p{1.5 cm}||p{6cm}| }
            \hline    
          Name\vspace{1 mm} &  Formula \vspace{1 mm} \\
           \hline
              Local defining function ($\hbar$) & $\hbar:\mathbb{R}^{3\times3} \rightarrow sym(3): X \rightarrow \hbar(X)=X^{\top}X-I_3$, where $det(X)=+1$ and $sym(3)$ is symmetric matrices of size 3 in the linear space and $\mathbb{R}^{3\times3}$ is the embedding space  \\ 
               \hline
               Tangent space ($\mathcal{T}_XSO(3)$) & $\mathcal{T}_X SO(3) = \{X\Omega \in \mathbb{R}^{3\times3} : X \in SO(3) \quad \& \quad \Omega \in Skew(3) \quad (\text{that means: } \Omega \in \mathbb{R}^{3\times3}: \Omega^{\top}=-\Omega ) \}$  \\
             \hline
             Vector Transport ($\mathcal{T}_{Y\leftarrow X}(\Vec{v})$) & $Y\Omega$, such that $\Vec{v} = X\Omega$

             \\
             \hline
                  Retraction ($\mathcal{R}_X(\Vec{v})$) & $qfactor(X+\Vec{v})$, where  $qfactor$ extracts the $Q$-$factor$ of the $QR$ decomposition with positive elements on the diagonal of $R$.  \\
           \hline                  
           \end{tabular}
           \label{tab:geometry for SO(3)}
       \end{table}    

\subsection{System Model}
This work considers tracking the position and orientation of a moving target equipped with IMU in an environment equipped with a measurement system (MS). The measurement system in this context means a system that uses the transmitted/ received signal for positioning. The positioning quantities are usually represented in two different but related coordinate systems: the body coordinate system (BCS) and the global coordinate system (GCS). To avoid confusion, the quantities in the BCS are denoted with a superscript $b$ in this script, while no superscript is used for those in GCS.

\begin{figure}[!ht]
    \centering
    \includegraphics[scale=0.15]{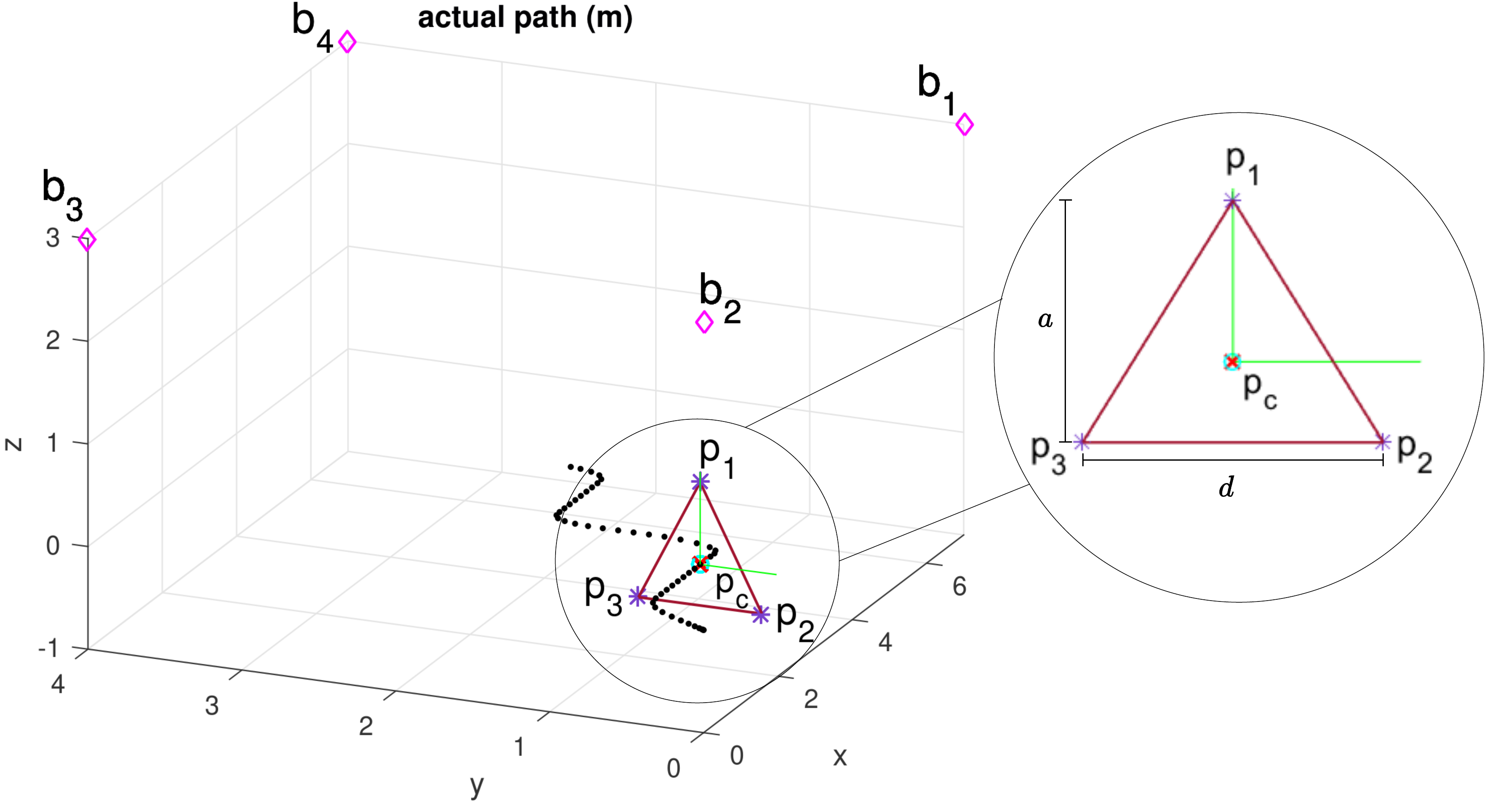}    \caption{System diagram shows the moving target with its components, where the dots line, brown triangle, purple asterisks, green axes, red cross, cyan circle, and magenta diamond refer to the moving path, isosceles triangle, transmitters, BCS, target centroid, IMU, and beacons, respectively.}
    \label{fig: SysMod}
\end{figure}

The considered MS consists of $M$ anchors (beacons) and three transmitters attached to the target. Swapping transmitters and receivers has no impact on the proposed algorithm. The position of the $i^{th}$ transmitter in the BCS and the $j^{th}$ anchor in GCS are known and denoted as $\mathbf{p}_{i}^{b}$ and $\mathbf{b}_{j}$, respectively, where both $\in \mathbb{R}^3$. The three transmitters are arranged to align the vertices of an isosceles triangle with a base of length $d$ and an altitude of length $a$, as depicted in Fig. \ref{fig: SysMod}.  To form an isosceles triangle, the transmitters' positions are formulated in terms of $a$ and $d$ as $\mathbf{p}^{b}_1=(0,0,\frac{2a}{3})$, $\mathbf{p}^{b}_2=(\frac{d}{2},0,\frac{-a}{3})$, and $\mathbf{p}^{b}_3=(\frac{-d}{2},0,\frac{-a}{3})$. Furthermore, the $s_{ij}$ denotes to the noisy range between $i^{th}$ transmitter and $j^{th}$ anchor such that $s=s_{true} + s_{n}$; where $s_n$ is an additive Gaussian noise with zero mean and $\sigma_{s}^{2}$ variance.

The IMU gives two quantities: the target centroid's acceleration in BCS ($\mathbf{a}^b_c \in \mathbb{R}^{3}$) and the angular velocity ($\mathbf{\omega} \in \mathbb{R}^3$). Both $\mathbf{a}^b_c$ and $\mathbf{\omega}$ form a column vector $\mathbf{u}\in \mathbb{R}^6$ such that $\mathbf{u} = [ \mathbf{\omega}^{\top} \quad \mathbf{a}^{b\top}_c ]^\top$, where $\mathbf{u}$ represents the input of our system.

The input $\mathbf{u}$ is a noisy quantity such that $\mathbf{u}= \mathbf{u}_{true} + \mathbf{u}_n$ where $\mathbf{u}_n$ is a noise process described as additive white Gaussian noise (AWGN) such that $\mathbf{u}_n  \sim \mathcal{N}(\mathbf{0}, \mathbf{Q})$. $\mathbf{Q}\in \mathbb{R}^{6\times6}$ is a diagonal matrix that represents the covariance matrix of the input. The first three diagonal elements represent the variance of the angular velocity $\sigma^2_{\omega}$ and the other three diagonal elements represent the variance of the acceleration $\sigma^2_{a}$. 

The position of the transmitters ($\mathbf{p}_1,\mathbf{p}_2,\mathbf{p}_3$) in GCS represents the output $\mathbf{y} \in \mathbb{R}^9$ of our system such that $\mathbf{y} = [ \mathbf{p}_1^{\top} \quad \mathbf{p}_2^{\top} \quad \mathbf{p}_3^{\top}]^\top$. This work utilizes two independent subsystems to estimate $\mathbf{y}$: the IMU system and MS system. To avoid confusion $\widehat{\mathbf{y}}_{imu}$ and ${\mathbf{y}}_{ms}$ refer to the estimated $\mathbf{y}$ based on the IMU and MS, respectively.  The ${\mathbf{y}}_{ms}$ is a noisy quantity such that ${\mathbf{y}}_{ms}=\mathbf{y}_{ms,true} + \mathbf{y}_{ms,n}$ where $\mathbf{y}_{ms,n} \sim \mathcal{N}(\mathbf{0},\mathbf{Z})$), and $\mathbf{Z}=\sigma^2_s I_9 \in \mathbb{R}^{9\times9}$ is the covariance matrix.

The ultimate purpose of this work is to provide an algorithm that estimates the position and orientation of the moving target subject to the mentioned circumstances by assuming a prior knowledge of the initial values of its position, velocity, and orientation in the global coordinate system. 

For simplicity, we assume here that the IMU, the origin of the BCS, the centroid of the target, and the centroid of the isosceles triangle coincide with each other. As a result, the position, velocity, and acceleration of the target are measured and estimated at its centroid, and they are denoted as $\mathbf{p}_c$, $\mathbf{v}_c$, and $\mathbf{a}_c$ respectively, all belonging to $\mathbb{R}^3$. Since the target's centroid coincides with the BCS's origin, $\mathbf{p}^{b}_c=(0,0,0)$. 

 Although we have access to some of the aforementioned positioning quantities in BCS solely, practically we are interested in them in the GCS. So, a matrix called rotation matrix $\mathbf{R} \in \mathbb{R}^{3\times 3}$, belonging to SO($3$) manifold, is used to transform the quantities from the BCS into the GCS and vice versa, as described in \cite{titterton2004strapdown}. The rotation matrix could be expressed as $\mathbf{R}= [R_1 \quad R_2 \quad R_3]$, where $R_i \in \mathbb{R}^3$ refers to the $i^{th}$ column of $\mathbf{R}$. Moreover, the element of $\mathbf{R}$ at $i^{th}$ row and $j^{th}$ column is denoted as $r_{ij}$.

\subsection{Problem Formulation}
\label{sec:prob.Formul}

This work adopts a state space model to formulate tracking the position and orientation of a moving target. This subsection is divided into two parts. The first part introduces the state space model in continuous time, while the second part focuses on discretizing the state space model. \\

\begin{enumerate}
    \item  \textbf{ State Space Model:} 
The state space model consists of two equations: a state equation and a measurement equation. 
The state equation of our model is described as follows. 

Let $\mathbf{\Theta} \in \mathbb{R}^9$ be a column vector with a column-major order of $\mathbf{R}$ (i.e. $\mathbf{\Theta}=[R_1^{\top} \quad R_2^{\top} \quad R_3^{\top}]^\top$). Also,  let $\mathbf{\Psi} \in \mathbb{R}^6$ be a column vector that contains the target position and velocity in GCS (i.e. $\mathbf{\Psi}=[\mathbf{p}_c^{\top} \quad \mathbf{v}_c^{\top} ]^\top$). Then, we define $\mathbf{X}$ as a state vector with dimension $L$ such that $\mathbf{X}= [\mathbf{\Theta}^{\top} \quad \mathbf{\Psi}^{\top}]^\top$ where  $\mathbf{X} \in \mathbb{R}^{15}$. From \cite[Sec 3.6]{titterton2004strapdown} and from the kinematic equations\cite{teodorescu2007mechanical}, it can be shown that the state equation of $\mathbf{X}$ can be represented as follows:
     \begin{equation}
         \dot{\mathbf{X}}= \mathbf{A}\mathbf{X}+ f(\mathbf{X})\mathbf{u}
        \label{eq:change of states with T}
     \end{equation}
     where
            \begin{equation}
                \mathbf{A}= \begin{bmatrix} \mathbf{0}_{9 \times 12} & \mathbf{0}_{9 \times 3} \\ \mathbf{0}_{3 \times 12} & \mathbf{I}_{3 \times 3} \\ \mathbf{0}_{3 \times 12} & \mathbf{0}_{3 \times 3} \end{bmatrix},    f(\mathbf{X})= \begin{bmatrix} \mathbf{\Pi}_{9 \times 3} & \mathbf{0}_{9 \times 3} \\ \mathbf{0}_{3 \times 3} & \mathbf{0}_{3 \times 3} \\ \mathbf{0}_{3 \times 3} & \mathbf{R}_{3 \times 3} \end{bmatrix}
            \end{equation}
        and where  the matrix $\mathbf{\Pi}$ is defined as: 
        \begin{equation}
            \mathbf{\Pi} = \begin{bmatrix}
                \mathbf{\Pi}_1 & \mathbf{\Pi}_2 & \mathbf{\Pi}_3
            \end{bmatrix}^\top
        \end{equation}
        such that:
        \begin{equation}
            \mathbf{\Pi}_1 = \begin{bmatrix}
                0 & 0 &0 \\-r_{13} &-r_{23} &-r_{33}  \\ r_{12} &  r_{22} &r_{32}
            \end{bmatrix},\end{equation} 
            \begin{equation}\mathbf{\Pi}_2 = \begin{bmatrix}r_{13} &r_{23} &r_{33} \\ 0 &0 &0 \\ -r_{11} &-r_{21} & -r_{31}
            \end{bmatrix},
        \end{equation} and 
            \begin{equation}
                \mathbf{\Pi}_3 = \begin{bmatrix}    -r_{12} &-r_{22} &-r_{32} \\
                r_{11} & r_{21} & r_{31} \\  0 & 0 & 0 \end{bmatrix}
            \end{equation}

    Now  we discuss the measurement equation of our model. 
    Since the $\mathbf{R}$ and $\mathbf{p}_c$ are known from the state equation, and since the transmitters' positions in BCS are fixed and also known, the output of the system $\mathbf{y}$ could be estimated based on the IMU readings as:
    \begin{align}
    \label{eq:yfromX}   
    \widehat{\mathbf{y}}_{imu}=h(\mathbf{X})= \begin{bmatrix} \mathbf{p}_c + \mathbf{R}\mathbf{p}^{b}_1  \\  \mathbf{p}_c + \mathbf{R}\mathbf{p}^{b}_2 \\  \mathbf{p}_c + \mathbf{R}\mathbf{p}^{b}_3 
         \end{bmatrix} 
    \end{align}
    In addition to estimating the output based on the IMU, KF needs to measure the output by utilizing the MS where several algorithms can be used. Two of these algorithms are the Gauss-Newton algorithm and the isosceles triangle manifold algorithm \cite[Sec 10.2]{alsharif2021manifold,dennis1996numerical}. \\
     
  \item \textbf{Discretization of State Space Model}

In order to derive the discrete-time model, certain assumptions are made about the discretized signals between the sampling instants. Here, we assume that the inputs remain constant, which is commonly referred to as the Zero-Order Hold (ZOH) assumption. 
Since each of $\mathbf{\Theta}$ and $\mathbf{\Psi}$ has a different characteristic, they are discretized separately. Then, these discretized equations are merged to form the overall discretized model.

       Since $\dot{\mathbf{\Theta}}_k\approx \frac{\mathbf{\Theta}_{k}-\mathbf{\Theta}_{k-1}}{T} \Longrightarrow \mathbf{\Theta}_{k}\approx\mathbf{\Theta}_{k-1} + T \dot{\mathbf{\Theta}}_k$. From $\eqref{eq:change of states with T}$, we can conclude that:
           \begin{align}
           {\dot{\Theta}}= \mathbf{\Pi}\mathbf{\omega} 
          \end{align}
          So, the discretized state space equation of $\mathbf{\Theta}$ will be as:
          \begin{align}     \widehat{\mathbf{\Theta}}_{k}&\approx\mathbf{\Theta}_{k-1} + T\mathbf{\Pi}\mathbf{\omega}
          \label{dis.theta}
          \end{align}

      From $\eqref{eq:change of states with T}$ also, we can conclude that:
      \begin{align}
      {\dot{\mathbf{\Psi}}}= \mathbf{\Upsilon} . \mathbf{\Psi} + \mathbf{B} .  \mathbf{a}^b_c            \end{align}
           where \begin{align}
               \mathbf{\Upsilon}= \begin{bmatrix} 
              {\mathbf{0}}_{3 \times 3} & {\mathbf{I}}_{3 \times 3} \\ {\mathbf{0}}_{3 \times 3} & {\mathbf{0}}_{3 \times 3}
               \end{bmatrix}, \quad
               \mathbf{B}= \begin{bmatrix} {\mathbf{0}}_{3 \times 3} \\ \mathbf{R}_{3 \times 3}
               \end{bmatrix}
           \end{align}
Since the state equation of $\mathbf{\Psi}$ is linear, it is convenient to use linear ZOH discretization equations. We need to find two matrices $\mathbf{\Upsilon}_D, \mathbf{B}_D$ describing the discrete-time system\cite[Sec 4.2]{chen1984linear}. These are given in terms of the continuous-time system matrices $\mathbf{\Upsilon}, \mathbf{B}$ and sampling time $T$ by:
\begin{equation}
    \mathbf{\Upsilon}_D = e^{\mathbf{\Upsilon}T},\quad \mathbf{B}_D = \int_0^T e^{\mathbf{\Upsilon}t}\mathbf{B}\, dt
\end{equation}
Applying these formulae we get:
$$
\mathbf{\Upsilon}_D = \begin{bmatrix}
\mathbf{I}_3 & T\mathbf{I}_3 \\ \mathbf{0}_3 & \mathbf{I}_3
\end{bmatrix}, \quad \mathbf{B}_D = \begin{bmatrix}
\frac{T^2}{2} \mathbf{R} \\ T\mathbf{R}
\end{bmatrix}
$$   
          \begin{align}
\widehat{\mathbf{\Psi}}_{k+1}&\approx\mathbf{\Upsilon}_D \mathbf{\Psi}_{k} + \mathbf{B}_D \mathbf{a}^b_c
\label{dis.psi}
             \end{align}
             Note that the discretization errors of (\ref{dis.theta}) and (\ref{dis.psi}) are proportion to $T$ and $T^{2}$ respectively.
             
      To define the discretized overall state space model, we combine the results of discretizing $\mathbf{\Theta}$ and $\mathbf{\Psi}$ as follows:
                \begin{align}
          \label{eq:time updat_MergedModel}
          \widehat{\mathbf{X}}_{k}&=\mathbf{F} \mathbf{X}_{k-1} + g(\mathbf{X}_{k-1})\mathbf{u}_{k-1}  
          \end{align}  
          \begin{align}
          \text{where}
          \label{EKF_F}
          \quad \mathbf{F}&= \begin{bmatrix} 
          \mathbf{I}_{9 \times9} & \mathbf{0}_{9 \times3} & \mathbf{0}_{9 \times3} \\ \mathbf{0}_{3 \times 9} & \mathbf{I}_{3 \times 3} & T\mathbf{I}_{3 \times 3} \\ \mathbf{0}_{3 \times 9} & \mathbf{0}_{3 \times3} & \mathbf{I}_{3 \times3}
          \end{bmatrix} \end{align} and \begin{align} \label{EKF_G} g(\mathbf{X}_{k-1}) = \begin{bmatrix} T\mathbf{\Pi}_{9 \times 3} & \mathbf{0}_{9 \times3} \\ \mathbf{0}_{3 \times 3} & \frac{T^2}{2}\mathbf{R} \\ \mathbf{0}_{3 \times 3} & T\mathbf{R}  \end{bmatrix}
             \end{align}   
\end{enumerate}

\section{PROPOSED ALGORITHMs}
\label{sec:proposed.Algor}

This section outlines the proposed algorithms for tracking moving targets. These algorithms are updated versions of the EKF and UKF, which leverage Riemannian optimization operators. For more details about EKF and UKF, see  \cite{sayed_2022}.

The rotation matrix belongs to the SO($3$) manifold, and the eigenvectors of its covariance matrix are assumed to be on the corresponding tangent space \cite{hauberg2013unscented}. If the noise is high, this assumption will not hold. In the estimation process and due to noise, the estimated rotation matrix is more likely to be on the tangent space of the SO(3) manifold. To address this, we use the retraction operator to project the rotation matrix from the tangent space into the SO(3) manifold. Hauberg \textit{et al.} \cite{hauberg2013unscented} proposed a method to update the corresponding covariance matrix as follows where the superscript ($\diamond$) indicates those quantities subjected to SO(3); and $\mathbf{P}^{\Theta}$ indicates the submatrix of $\mathbf{P}$ associated with $\mathbf{\Theta}$. First, the corresponding covariance matrix from the previous time instance ($\mathbf{P}^{\diamond\Theta}_{k-1}$) is decomposed into its eigenvectors ($\mathbf{V}$) and eigenvalues ($\Lambda$). Subsequently, the parallel transport operation is applied to shift the eigenvectors from the tangent space of the original rotation matrix to the tangent space of the updated rotation matrix. The resulting transported eigenvectors, denoted by $\mathbf{V}^{\diamond}$, are then recombined with the original eigenvalues to form the new covariance matrix as follows:
 \begin{equation}
     \label{eq:CovUpdate}
     \mathbf{P}^{\diamond\Theta}_{k} = \mathbf{V}^{\diamond}\mathbf{\Lambda}\mathbf{V}^{\diamond\top}
 \end{equation}
\renewcommand{\algorithmicrequire}{\textbf{Input:}}
\renewcommand{\algorithmicensure}{\textbf{Output:}}
    \begin{algorithm}[!h]
    \caption{: Proposed EKFRie algorithm}
        \label{alg:EKFRie}

    \begin{algorithmic}[1]
    \small
        \REQUIRE 
     $ \mathbf{\widehat{X}^{\diamond}}_{k-1|k-1}, \mathbf{y}_{ms,k}, \mathbf{a}^b_{imu,k-1}, \mathbf{\omega}_{k-1}, \mathbf{P}^{\diamond}_{k-1|k-1}$, \\ $\text{and } \mathbf{p}^{b}_{1,2,3}$.  
        \ENSURE
     $\mathbf{\widehat{X}}^{\diamond}_{k|k}\quad $ and $\quad \mathbf{P}^{\diamond}_{k|k}$.
     \\ \hspace{-1.5em}{\textbf{Steps:}}\\
          \STATE $k_1  =  MS_{DataRate}$\\
      \hspace{-1.5em} \textbf{repeat for} $k > 0$
      \STATE $\textbf{a}^{b}_{c,k-1} \gets \textbf{a}^{b}_{imu,k-1}$ \& $\sigma_{a,k-1}^{2} \gets \sigma_{aimu,k-1}^{2}$ by using \eqref{eq:a_imu2a_c} \& \eqref{eq:VAR_aimu2Var_ac}\\
     \STATE Find $ g(\mathbf{X}^{\diamond}_{k-1|k-1})$ by using  \eqref{EKF_G}\\
     \STATE Find $\widehat{\mathbf{X}}_{k|k-1}$ by using \eqref{eq:time updat_MergedModel} and \eqref{EKF_F}
     \STATE \footnotesize $\mathbf{P}_{k|k-1}\gets  \mathbf{F} \mathbf{P}^{\diamond}_{k-1|k-1}\mathbf{F}^{\top}+g(\mathbf{X}^{\diamond}_{k-1|k-1})\mathbf{Q}_{k} g(\mathbf{X}^{\diamond}_{k-1|k-1})^{\top}$
      \STATE \small $\widehat{\mathbf{\Theta}}^{\diamond}_{k|k-1}\gets \mathcal{R}_{\mathbf{\Theta}^{\diamond}_{k-1|k-1}}(T\mathbf{\Pi}_{k-1|k-1}\mathbf{\omega}_{k-1})$   
    \STATE $\mathbf{P}^{\diamond\Theta}_{k|k-1} \gets \mathbf{P}^{\diamond\Theta}_{k-1|k-1}\text{ by using }$ \eqref{eq:CovUpdate}\\
        \STATE $\mathbf{\widehat{\mathbf{X}}}^{\diamond}_{k|k-1} \gets \begin{bmatrix}
            \mathbf{\widehat{\Theta}}^{\diamond}_{k|k-1} & \mathbf{\widehat{\Psi}}_{k|k-1}
                 \end{bmatrix}^{\top}$ \\
                 
                  \IF{$\text{Mod}(k_1, MS_{DataRate}) = 0$}
     \STATE Find $\widehat{\mathbf{y}}_{imu,k|k-1}$ by using \eqref{eq:yfromX}
         \STATE $\mathbf{H}_k \gets \frac{\partial h(\mathbf{X})}{\partial \mathbf{X}}|_{\mathbf{X}=\widehat{\mathbf{X}}^\diamond_{k|k-1}}$
        \STATE $\mathbf{K}_{f,k} \gets \mathbf{P}^{\diamond}_{k|k-1}\mathbf{H}^\top_{k} (\mathbf{H}_{k} \mathbf{P}^{\diamond}_{k|k-1} \mathbf{H}^\top_{k} + \mathbf{Z}_k )^{-1}$
        \STATE $\widehat{\mathbf{X}}_{k|k} \gets \widehat{\mathbf{X}}^{\diamond}_{k|k-1}+\mathbf{K}_{f,k}(\mathbf{y}_{ms,k} -\widehat{\mathbf{y}}_{imu,k|k-1})$
        \STATE $\mathbf{P}_{k|k} \gets (\mathbf{I}-\mathbf{K}_{f,k}\mathbf{H}_k)\mathbf{P}^{\diamond}_{k|k-1}$
    \STATE  $\widehat{\mathbf{\Theta}}^{\diamond}_{k|k} \gets \mathcal{R}_{\mathbf{\Theta}^{\diamond}_{k|k-1}}$ (the corresponding part of  \\ $\mathbf{K}_{f,k}(\mathbf{y}_{ms,k} -\widehat{\mathbf{y}}_{imu,k|k-1}) \text{ to }\mathbf{\Theta})$
   \STATE $\mathbf{P}^{\diamond\Theta}_{k|k} \gets \mathbf{P}_{k|k-1}^{\diamond\Theta}$ by using \eqref{eq:CovUpdate}\\ 
        \STATE $\mathbf{\widehat{\mathbf{X}}}^{\diamond}_{k|k} \gets \begin{bmatrix}
            \mathbf{\widehat{\Theta}}^{\diamond}_{k|k} & \mathbf{\widehat{\Psi}}_{k|k}
      \end{bmatrix}^{\top}$
           \ENDIF
          \\ \STATE $k_{1} = k_{1}+ 1$
     \\  \hspace{-1.5em}{\textbf{end}}.
    \end{algorithmic}
        \label{alg:EKF_hammam}
    \end{algorithm}

 Although we use similar ideas to Hauberg, we are more interested in the specific
application of indoor navigation. Hauberg applies his ideas to articulated tracking. In our work the SO(3) manifold is
quite appropriate for the indoor navigation with multiple receivers case, whilst the angles manifold is appropriate for the
articulated tracking example. Also, as parallel transport is computationally expensive \cite{boumal2020introduction, AbseOptimizationAlgorithmsl}, we replace it with the vector transport operation. Hauberg uses the exponential map with UKF, whilst we relax this to using a retraction with both
UKF and EKF demonstrate that the retraction can also work. The retraction and vector transport are used after the time update and measurement update in our work.
Finally, the case in Hauberg contains measurements and time updates at the same rate, while in our case the measurements occur at a slower rate.

 One of the UKF's challenges is losing the covariance matrix's positive definiteness property. This happens in this work, particularly after applying vector transport. As a result of that, the Cholesky factorization fails. This challenge of the UKF is known in the literature as stated in \cite[Sec. 30.8]{sayed_2022} and \cite{emami2014particle}. To overcome this issue, this work replaces the transported covariance matrix with its nearest positive definite matrix by utilizing the nearestSPD Matlab function \cite{nearestSPD}.  
 
As outlined in table \ref{tab:geometry for SO(3)}, the retraction operator for the SO(3) manifold can be expressed as a Q-factor of the sum of a matrix belonging to SO(3) manifold and the related tangent vector. Furthermore, Section \ref{sec:prob.Formul} introduces the time update equation for $\boldsymbol{\Theta}$, given by \eqref{dis.theta}.
We aim to demonstrate that the right-hand side of \eqref{dis.theta} can be restructured to match the structure of the $qfactor(.)$ function argument, namely, a matrix belonging to the SO(3) manifold plus a tangent vector of the SO(3) manifold. This allows us to directly apply the retraction operator by simply putting the restructured right-hand side of \eqref{dis.theta} as an argument of the $qfactor(.)$ function.

To show that, let us convert \eqref{dis.theta} from its vector form into matrix form as follows:
\begin{equation}
\label{eq.matrixFormofTU}
     \widehat{\mathbf{R}}_{k}\approx \mathbf{R}_{k-1}+T\mathbf{R}\boldsymbol{\Omega}
\end{equation}
where
      \begin{equation}
      \label{eq.omeg}
        \boldsymbol{\Omega} = \begin{bmatrix}
         0         & -\omega_z  & \omega_y \\
         \omega_z  &    0       & -\omega_x \\
         -\omega_y &   \omega_x &      0    \\
         \end{bmatrix}
      \end{equation}

 Eq. \eqref{dis.theta}  and \eqref{eq.matrixFormofTU} are equivalent to each other. The first term of \eqref{eq.matrixFormofTU}  is clearly on the SO(3) manifold. The $\boldsymbol{\Omega}$, in the second term, is a skew symmetric matrix (meaning $\boldsymbol{\Omega}^\top = -\boldsymbol{\Omega}$), and its scaling is also skew symmetric. Since the tangent space of SO(3) is defined in Table \ref{tab:geometry for SO(3)} as the product of a matrix on the SO(3) manifold and a skew symmetric matrix, the second term of  \eqref{eq.matrixFormofTU} is in the tangent space of the SO(3) manifold. Therefore, it has been shown that the right-hand side of \eqref{dis.theta} can be reformulated to match the structure of the argument of the $qfactor(.)$ function.

\renewcommand{\algorithmicrequire}{\textbf{Input:}}
\renewcommand{\algorithmicensure}{\textbf{Output:}}
    \begin{algorithm}[ht!]
    \small
    \caption{:Proposed UKFRie algorithm}
        \label{alg:UKFRie}
    \begin{algorithmic}[1]
        \REQUIRE 
     $ \mathbf{\widehat{X}^{\diamond}}_{k-1|k-1}, \mathbf{y}_{ms,k}, \mathbf{a}^b_{imu,k-1}, \mathbf{\omega}_{k-1}, \mathbf{P}^{\diamond}_{k-1|k-1}$, $\text{ and } \mathbf{p}^{b}_{1,2,3}$.  
        \ENSURE
     $\mathbf{\widehat{X}}^{\diamond}_{k|k} $ and $ \mathbf{P}^{\diamond}_{k|k}$.
     \\ \hspace{-1.5em}{\textbf{Steps:}}\\
          \STATE $k_1  = MS_{DataRate}$\\
      \hspace{-1.5em} \textbf{repeat for} $k > 0$:
       \STATE $\textbf{a}^{b}_{c,k-1} \gets \textbf{a}^{b}_{imu,k-1}$ \& $\sigma_{a,k-1}^{2} \gets \sigma_{aimu,k-1}^{2}$ by using \eqref{eq:a_imu2a_c} \& \eqref{eq:VAR_aimu2Var_ac}\\
      \STATE Finding sigma points ($\mathbf{\chi}_{\ell}$) and weights ($\xi_{\ell}$) of $\mathbf{\widehat{X}}^{\diamond}_{k-1|k-1}$ based on \cite{sayed_2022} with $\iota = 1$, $\alpha= 0.001$, and $\ell=1,2,\dots,2L$\\
       \STATE Finding $ g(\mathbf{\chi}_{\ell})$ by using \eqref{EKF_G} and   $\mathbf{X}_{k,\ell}$  by using \eqref{eq:time updat_MergedModel} and \eqref{EKF_F}\\       
         \STATE $\widehat{\mathbf{X}}_{k|k-1} \gets \sum^{2L}_{\ell=0} \xi_\ell \mathbf{X}_{k,\ell}$ and $\mathbf{P}_{k|k-1} \gets \sum^{2L}_{\ell=0} \xi_\ell (\mathbf{X}_{k,l} - \widehat{\mathbf{X}}_{k|k-1})(\mathbf{X}_{k,l} -\widehat{\mathbf{X}}_{k|k-1})^\top +\xi^2_\ell g(\mathbf{\chi}_{\ell}) \mathbf{Q}_k g(\mathbf{\chi}_{\ell})^{\top}$\\ 
            \STATE $\widehat{\boldsymbol{\Theta}}^{\diamond}_{k|k-1} \gets \mathcal{R}_{\boldsymbol{\Theta}^{\diamond}_{k-1|k-1}}( \sum^{2L}_{\ell=0} \xi_{\ell}T\boldsymbol{\Pi}_{\ell}\mathbf{\omega}_{k-1})$  where $ \boldsymbol{\Pi}_{\ell} \text{ corresponding to the } \chi_{\ell}$

    \STATE  $ \mathbf{P}^{\diamond\Theta}_{k|k-1} \gets nearestSPD(\mathbf{V}^{\diamond}\boldsymbol{\Lambda}\mathbf{V}^{\diamond\top}) \gets \mathbf{V}^{\diamond} \gets \mathcal{T}_{\boldsymbol{\Theta}^{\diamond}_{k|k-1} \leftarrow \boldsymbol{\Theta}^{\diamond}_{k-1|k-1}}(\mathbf{V}) \gets \mathbf{V}\boldsymbol{\Lambda}\mathbf{V}^{\top} \gets eig(\mathbf{P}_{k-1|k-1}^{\diamond\Theta})$

        \STATE $\mathbf{\widehat{\mathbf{X}}}^{\diamond}_{k|k-1} \gets \begin{bmatrix}
            \boldsymbol{\widehat{\Theta}}^{\diamond}_{k|k-1} & \boldsymbol{\widehat{\Psi}}_{k|k-1}
                 \end{bmatrix}^{\top}$ \\
         \IF{$ \text{Mod}(k_1, MS_{DataRate}=0$)}
          \STATE Finding sigma points of  $\mathbf{\widehat{\mathbf{X}}}^{\diamond}_{k|k-1}$ and use them in \eqref{eq:yfromX} to find $\mathbf{\widehat{y}}_{k,\ell}$ 
          
      \STATE $\widehat{\mathbf{y}}_{imu,k|k-1} \gets \sum^{2L}_{\ell=0} \xi_\ell \mathbf{\widehat{y}}_{k,\ell}$\\
      \STATE  $\delta_{e,k} \gets \sum^{2L}_{\ell=0} \xi_\ell (\mathbf{\widehat{y}}_{k,\ell} -\widehat{\mathbf{y}}_{imu,k|k-1})(\mathbf{\widehat{y}}_{k,\ell}-\widehat{\mathbf{y}}_{imu,k|k-1})^\top + \mathbf{Z}_k$\\
      \STATE $\delta_{xe,k} \gets \sum^{2L}_{\ell=0} \xi_\ell (\mathbf{X}_{k,\ell} -\widehat{\mathbf{X}}_{k|k-1})(\mathbf{\widehat{y}}_{k,\ell} -\widehat{\mathbf{y}}_{imu,k|k-1})^\top$\\
       \STATE $\mathbf{K}_{f,k} \gets \delta_{xe,k}\delta_{e,k}^{-1}$\\
        \STATE $\widehat{\mathbf{X}}_{k|k} \gets \mathbf{X}^{\diamond}_{k|k-1}+\mathbf{K}_{f,k}(\mathbf{y}_{ms,k} -\widehat{\mathbf{y}}_{imu,k|k-1})$ and $\mathbf{P}_{k|k} \gets \mathbf{P}^{\diamond}_{k|k-1} - \mathbf{K}_{f,k}\delta_{e,k}\mathbf{K}_{f,k}^{\top}$\\
      
      \STATE  $\widehat{\boldsymbol{\Theta}}^{\diamond}_{k|k} \gets \mathcal{R}_{\boldsymbol{\Theta}^{\diamond}_{k|k-1}}(\text{the corresponding part of }  \mathbf{K}_{f,k}(\mathbf{y}_{ms,k} -\widehat{\mathbf{y}}_{imu,k|k-1}) \text{to }\boldsymbol{\Theta})$
    \STATE $\mathbf{P}^{\diamond\Theta}_{k|k} \gets nearestSPD(\mathbf{V}^{\diamond}\boldsymbol{\Lambda}\mathbf{V}^{\diamond\top}) \gets \mathbf{V}^{\diamond} \gets \mathcal{T}_{\boldsymbol{\Theta}^{\diamond}_{k|k} \leftarrow \boldsymbol{\Theta}^{\diamond}_{k|k-1}}(\mathbf{V}) \gets \mathbf{V}\boldsymbol{\Lambda}\mathbf{V}^{\top} \gets eig(\mathbf{P}_{k|k-1}^{\diamond\Theta})$
    
        \STATE $\mathbf{\widehat{\mathbf{X}}}^{\diamond}_{k|k} \gets \begin{bmatrix}
            \boldsymbol{\widehat{\Theta}}^{\diamond}_{k|k} & \boldsymbol{\widehat{\Psi}}_{k|k}
      \end{bmatrix}^{\top}$

           \ENDIF
           \STATE $k_1 = k_1 + 1$
\\ \hspace{-1.5em}{\textbf{end}}.
           \end{algorithmic}
               \label{alg:UKF_hammam}
     \end{algorithm}

\begin{figure*}[hbt!]
\includegraphics[width=\textwidth,height=5cm]{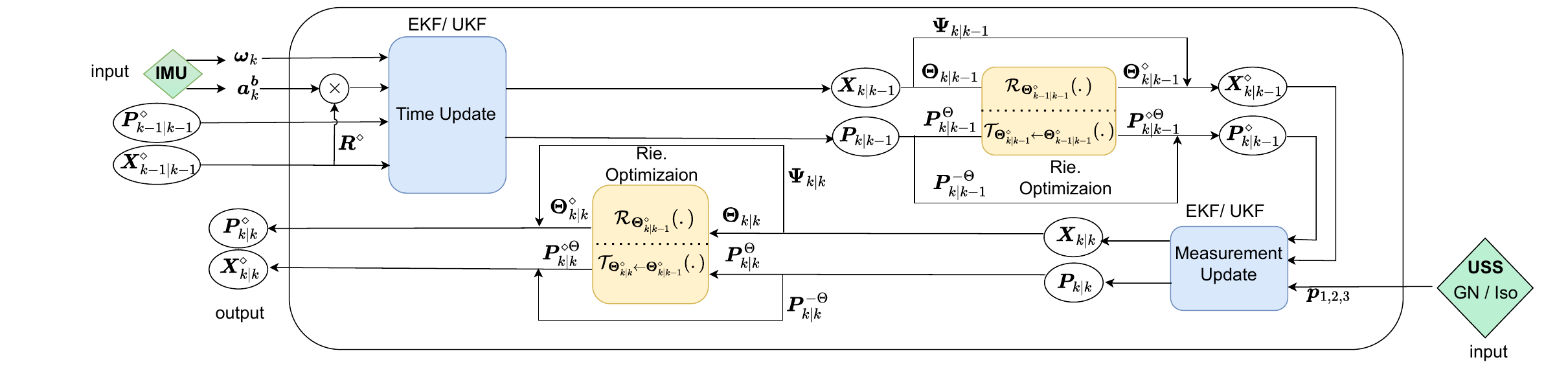}
    \caption{ Descriptive diagram of the proposed algorithms}
    \label{fig:Algor.Diag}
\end{figure*}

The overall mechanism of the algorithms is illustrated in Fig.\ref{fig:Algor.Diag} where GN and Iso refer to the Gauss-Newton algorithm and isosceles triangle manifold algorithm, respectively, and $\mathbb{P}^{-\Theta}$ refers to the part of $\mathbb{P}$ corresponding to elements other than the rotation matrix. The proposed algorithms, EKF with Riemannian (EKFRie) and UKF with Riemannian (UKFRie), are presented in Algorithms \ref{alg:EKFRie} and \ref{alg:UKFRie}, respectively.

In the state covariance matrix update in Algorithm.\ref{alg:EKFRie} (step.5), the author treats $g(\mathbf{X}^{\diamond}_{k-1|k-1})$ as a constant for simplicity. This is valid because the value of $\mathbf{P}_{k|k-1}$ depends mainly on the first term since its norm is much larger than the norm of the second term (approximately 40 times larger). In the Algorithms \ref{alg:EKF_hammam} and \ref{alg:UKF_hammam}, $\mathbf{P}^{\diamond}$ refers to $\mathbf{P}$ after replacing $\mathbf{P}^{\Theta}$ by $\mathbf{P}^{\diamond\Theta}$.

Due to engineering and manufacturing constraints, it is sometimes not possible to place the IMU at the centroid. To use the
models developed above we need to transform the IMU readings to the centroid. This affects the covariance of these readings
and how they are integrated into the Kalman Filter. The angular velocity needs no transformation as the angular velocity is
the same throughout rigid bodies. For generality, these pseudo codes consider this scenario , i.e. the IMU and the BCS's centroid don't coincide. The variables $\textbf{a}^{b}_{imu}$ and $\sigma_{aimu}^{2}$ refer to the acceleration and its variance at the IMU, respectively, and \eqref{eq:a_imu2a_c} and \eqref{eq:VAR_aimu2Var_ac} in appendix \ref{sec:Derivation} are used to transfer them into the BCS's centroid.

The proposed algorithms comprise two parts: the time update part and the measurement update part. The time update part, represented by the outer loop, handles the estimation of the target position and orientation based on IMU readings. On the other hand, the measurement update part, represented by the inner loop, focuses on correcting error drift using the MS's readings.

\section{Results and Discussion}
\label{sec:ResulandDiss}
In this section, we evaluate the proposed algorithm through simulations and experiments. Simulations allow us to conveniently vary the various noise parameters, while experiments ensure that the models are realistic. 

In simulations, we compare three versions of the two main categories (EKF and UKF). Consider the EKF category the three algorithms are: (1) EKF just the conventional EKF (2) EKFIso the conventional EKF but with position measurements coming from Riemannian Isosceles manifold calculations as in \cite{alsharif2021manifold} (3) EKFRie the proposed algorithm in this work. The metrics for comparison are the cumulative distribution function (CDF) and the RMSE where RMSE is calculated as the sum over the three axes and averaged over time.

\subsection{Simulation Setup}
The simulations were conducted using a MATLAB program running on a $64-bit$ computer with $8GB$ of RAM, powered by an Intel(R) Core(TM) i7-1165G7 CPU ($2.80 GHz$). All simulations were run for $3000$ iterations. The IMU considered for the simulations is the MPU-9250 due to its small size, low cost, and low complexity \cite{MPU9250SpecificationDatasheet}. It includes a $3-$axis gyroscope, $3-$axis accelerometer, and $3-$axis magnetometer. The MPU-9250 has a gyroscope update  rate range of $4-8000 Hz$ and an accelerometer update rate range of $4-4000 Hz$. In our simulations, the gyroscope and accelerometer update rates were fixed to $10 Hz$ to reduce simulation time, unless otherwise stated.

IMU noise characteristics were incorporated into the simulations based on the MPU-9250 data sheet \cite{MPU9250SpecificationDatasheet}. Gyroscope noise was modelled with a variance of $5 \times 10^{-4} degree/s$, and accelerometer noise was modelled with a variance of $0.43 \times 10^{-3} m/s^2$. These variances were calculated based on the gyroscope's rate noise spectral density ($0.01 degree/s/sqrt(Hz)$) and the accelerometer's noise power spectral density ($300 \mu g/sqrt(Hz)$), as provided in the data sheet. The variances were calculated using the following formulas: $\sigma^{2}_{a} = (300 \times 9.8 \times 10^{-6})^2 \times (f_s / 2)$ and $\sigma^{2}_{\omega} = (0.01)^{2}  \times (f_s/2)$, where $f_s$ is the updating rates.

The accuracy of measurement systems can vary depending on the type of sensor used. For example, Cricket RF-ultrasonic sensors \cite{Cricket} provide $1-3 $cm position precision, Marvelmind sensors \cite{US_StarterSetSuper_MP_3D}  provide accuracy of $\pm2 $cm or better, and UWB DW1000 sensors \cite{DW1000} provide around $10 $cm position precision. Ultrasonic systems generally provide accuracy within the range $3-100 $cm \cite{yasir2015indoor}. Therefore, this work studies the effect of MS standard deviation values within the range $1 $cm to $1 $m. The measurement system update rate was fixed at $1 Hz$ which is lower than the IMU's update rate to be consistent with real-world applications. The three receivers are in an isosceles triangle configuration with a base and altitude of $10$cm and $30$cm, respectively.

\begin{figure}[!ht]
        \begin{subfigure}[b]{0.2\textwidth}
             \hspace{-0.25 cm}\includegraphics[scale = 0.35]{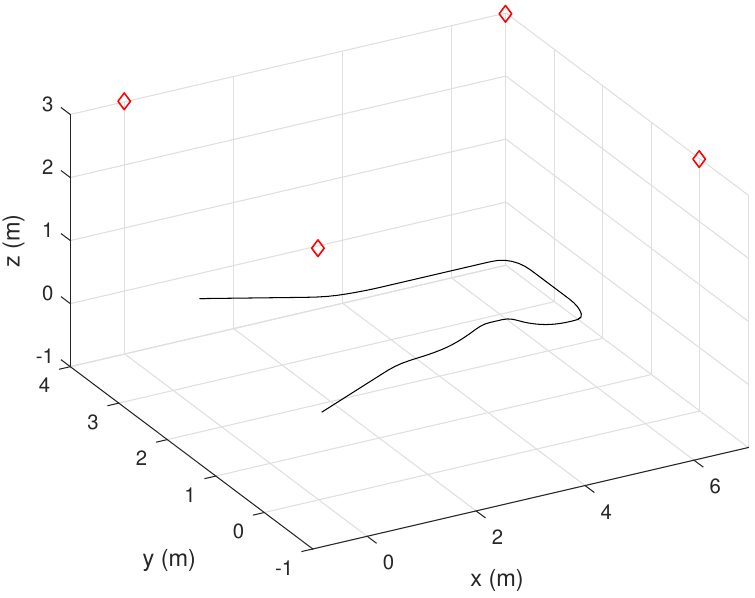}
            \caption[]%
            {{\small U-path}}    
        \end{subfigure}
        \hfill
        \begin{subfigure}[b]{0.2\textwidth}  
            \hspace{-0.9 cm}\includegraphics[scale = 0.35]{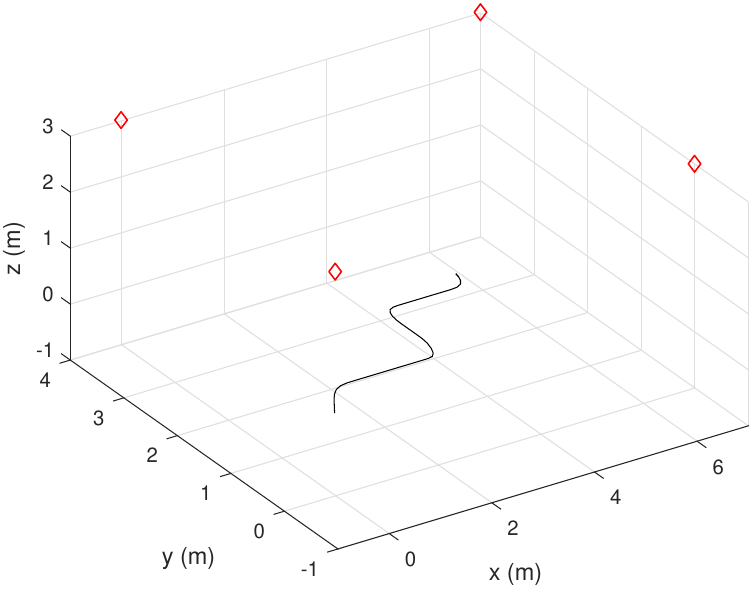}
            \caption[]%
            {{\small Zigzag-path}}    
        \end{subfigure}
            \vskip\baselineskip
        \begin{subfigure}[b]{0.2\textwidth}   
            \hspace{-0.5 cm}\includegraphics[scale = 0.35]{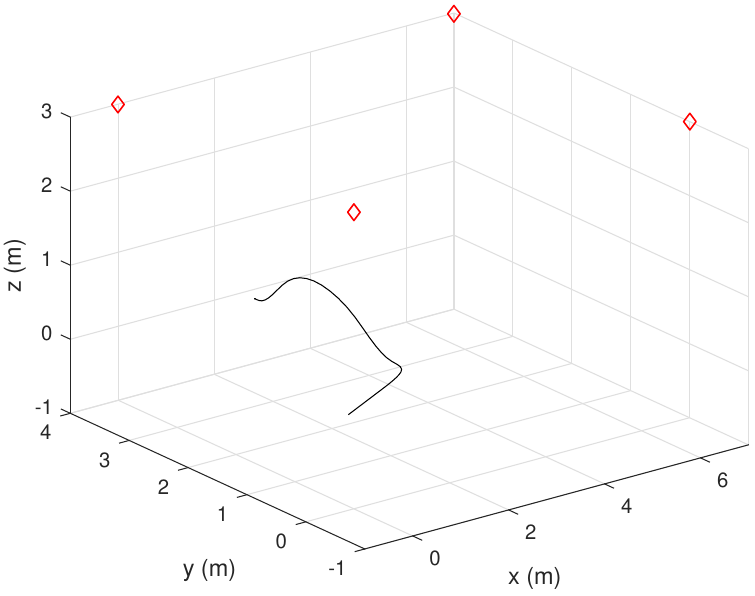}
            \caption[]%
            {{\small Bridge-path}}    
        \end{subfigure}
        \hfill
        \begin{subfigure}[b]{0.22\textwidth}   
            \hspace{-1.5 em}\includegraphics[scale = 0.32]{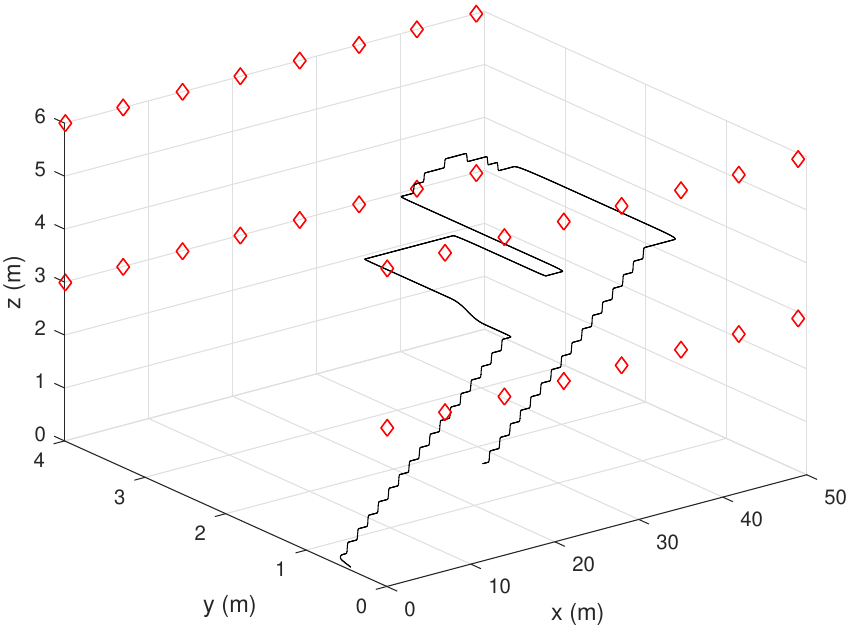}
            \caption[]%
            {{\small Stair-path}}    
        \end{subfigure}
        \caption[ ]
        {{\small The true four paths under consideration where the red diamonds refer to the beacons}} 
        \label{fig:true_paths}
    \end{figure}

The simulation scenarios we consider are (a) Static Scenario (b) Four dynamic scenarios: (1) U-path (2) Zigzag path (3) Bridge path (4) Stair path. These scenarios are shown in Fig. \ref{fig:true_paths}. The base stations are shown as red diamonds. These paths take $10$, $5$, $5$, and $85$ seconds, respectively, with a maximum velocity of less than $2 m/s$ to simulate human and robot \cite{PAPCUN2012533} motion. Additionally, this work examines the performance of the proposed algorithms under both static and dynamic conditions. 

\subsection{Simulation Results and Discussion} 
\begin{figure}[!ht]
        \centering
        \begin{subfigure}[b]{0.4\textwidth}
            \centering
            \includegraphics[width=\textwidth]{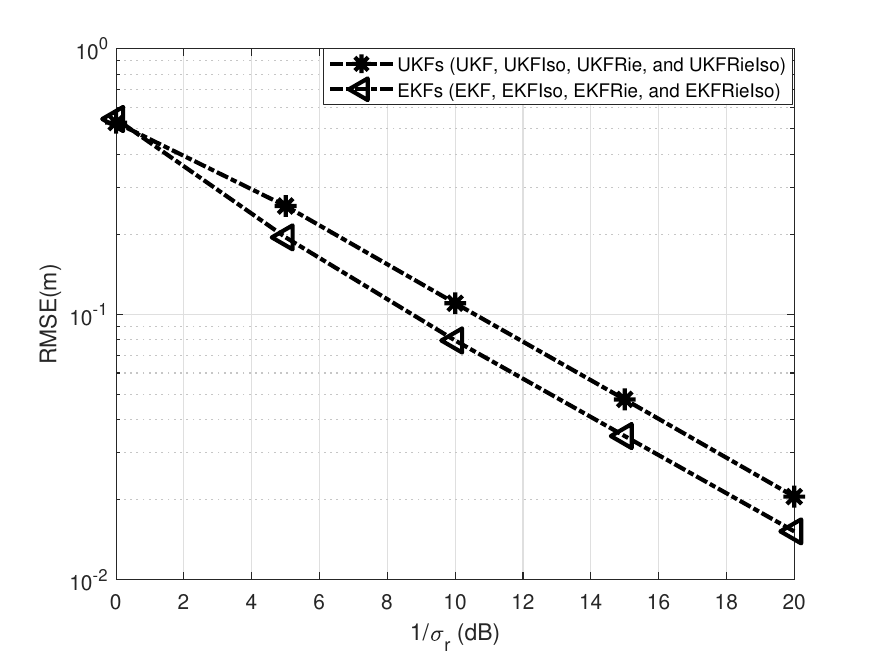}
            \caption[]%
            {{\small Position}}    
            \label{fig:RMSEvssigma_r_path5_a}
        \end{subfigure}
        \hfill
        \begin{subfigure}[b]{0.36\textwidth}  
            \centering 
            \includegraphics[width=\textwidth]{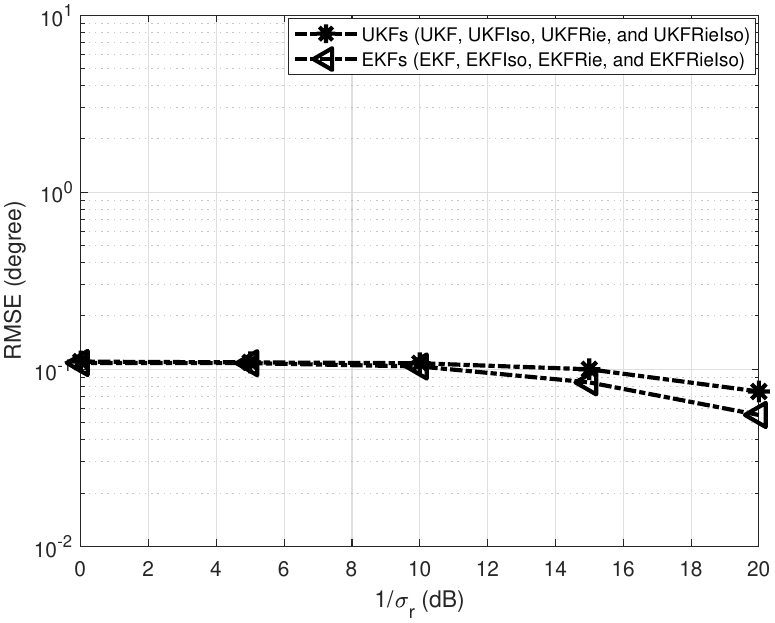}
            \caption[]%
            {{\small Orientation}}    
        \end{subfigure}
                    \caption[]%
        {{\small RMSE vs  of $1/ \sigma_r$ (dB) for  the static case }} 
          \label{fig:RMSEvssigma_r_path5}
      \end{figure}

In this section, we evaluate the proposed algorithms under static and dynamic scenarios. We employed the previously described model to simulate a static case with the acceleration, linear velocity, and angular velocity set to zero. Fig. \ref{fig:RMSEvssigma_r_path5} illustrates the RMSE for the estimated position and orientation plotted against the reciprocal of different values of the measurement system's standard deviation ($\sigma_r$). We use the dB unit because 
$1/\sigma_r$ closely resembles the mathematical form of signal-to-noise ratio (SNR). 
Fig. \ref{fig:RMSEvssigma_r_path5_a} shows that the position's RMSE of the EKFs and UKFs decrease linearly as $1/\sigma_r$ increases.

Furthermore, Fig. \ref{fig:RMSEvssigma_r_path5} shows that the performance of all EKFs (EKF, EKFIso, EKFRie, and EKFRieIso) are aligned closely, as the performance of all UKFs (UKF, UKFIso, UKFRie, and UKFRieIso). 

That is because the angular velocity is zero, which implies that the orientation matrix remains constant in the static scenario. Consequently, EKF and UKF perform better in this case, rendering the exploitation of the SO(3) structure of the rotation matrix redundant. This observation suggests that Riemannian optimization is practically useless in static scenarios.


         \begin{figure}[!ht]
        \centering
        \begin{subfigure}[b]{0.4\textwidth}
            \centering
            \includegraphics[width=\textwidth]{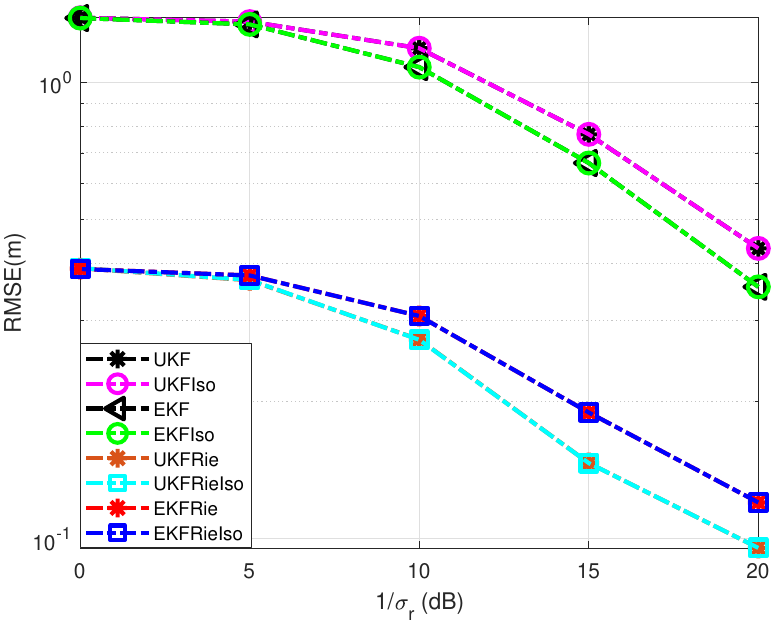}
            \caption[]%
            {{\small U-path}}    
            \label{RMSEvssigma_r_pos_path1}
        \end{subfigure}
        \hfill
        \begin{subfigure}[b]{0.4\textwidth}  
            \centering 
            \includegraphics[width=\textwidth]{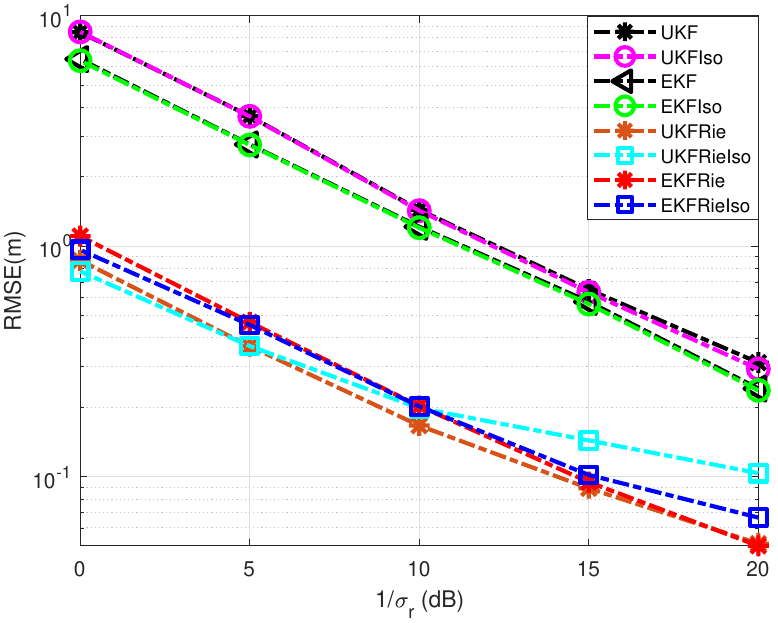}
            \caption[]%
            {{\small Stair-path}}    
            \label{RMSEvssigma_r_pos_path4}
        \end{subfigure}
        \caption[ ]
        {{\small Positions' RMSE vs different values of $1/ \sigma_r$ (dB) }} 
        \label{fig:RMSEvssigma_r_pos}
    \end{figure}
    
Regarding the dynamic scenario, the proposed algorithms consistently surpass the conventional EKF and UKF in terms of position and orientation across all paths and noise levels $\sigma_r$.  For U-path and $1/\sigma_r=20$ dB, the RMSE for the EKF and EKFRie were $0.36 $m and $0.12 $m, respectively, and for the UKF and UKFRie were $0.43 $m and $0.10 $m, respectively. For Stair-path and $1/ \sigma_r=20$ dB, the RMSE for the EKF and EKFRie were $0.24 $m and $0.05 $m, respectively, and for the UKF and UKFRie were $0.29 $m and $0.05 $m, respectively.

\begin{figure}[!ht]
        \centering
        \begin{subfigure}[b]{0.4\textwidth}
            \centering
            \includegraphics[width=\textwidth]{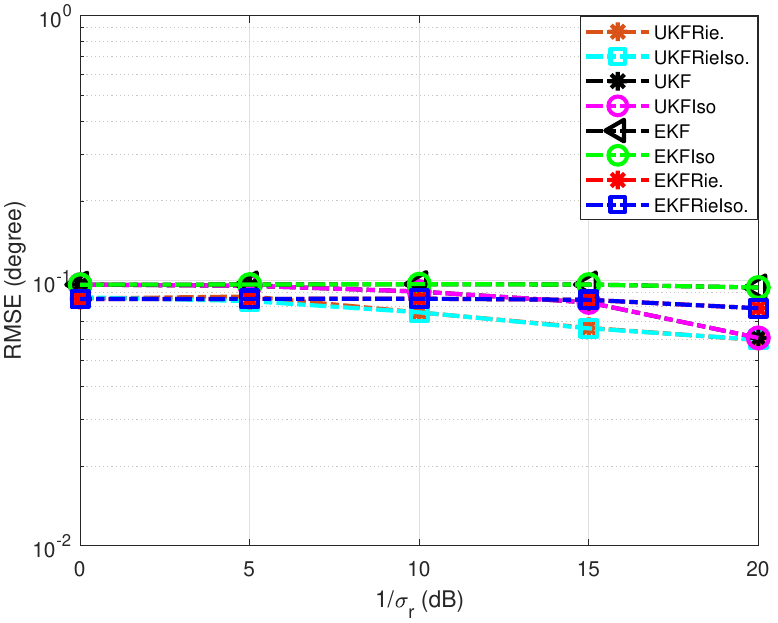}
            \caption[]%
            {{\small U-path}}    
        \end{subfigure}
        \hfill
        \begin{subfigure}[b]{0.4\textwidth}  
            \centering 
            \includegraphics[width=\textwidth]{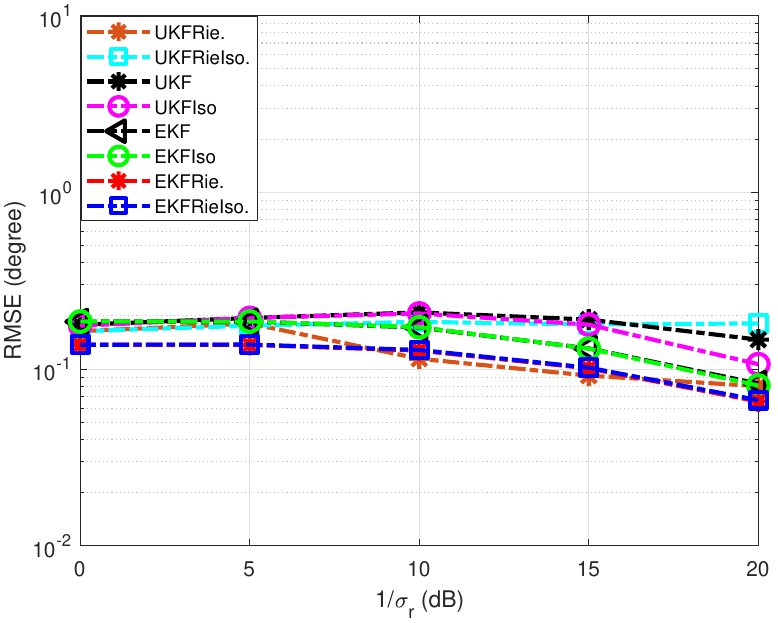}
            \caption[]%
            {{\small Stair-path}}    
            \label{fig:RMSEvssigma_r_theta_b}
        \end{subfigure}
        \caption[ ]
        {{\small Orientation's RMSE vs different values of $1/ \sigma_r$ (dB) }} 
        \label{fig:RMSEvssigma_r_theta}
    \end{figure}

 Remarkably, incorporating Riemannian optimization leads to a noticeable reduction in RMSE. Fig. \ref{fig:RMSEvssigma_r_pos} and \ref{fig:RMSEvssigma_r_theta} illustrate the RMSE of the estimated position and orientation versus $1/\sigma_r$ for the first and fourth paths, respectively, due to space limitations. In the legend, ``(Iso)" denotes the use of the isosceles triangle manifold algorithm during the measurement stage to estimate transmitters' positions, while the Gauss-Newton algorithm was employed for the others.

Using Riemannian only on measurements does not produce an advantage. Although \cite{alsharif2021manifold} shows that the position error of the measurement can be reduced (and our simulations have confirmed this), when this is used in the EKFs and UKFs no advantage appears. This strongly motivates this work where the Riemannian tools are employed after every step of the Kalman filter rather than only to process the range measurements. This is explained by noting that the improvement given by the Isosceles manifold increases as the range noise increases. However, the contribution of the measurement system to the Kalman filter will decrease with range noise. These effects seem to counteract each other leading to no overall gain in accuracy.

One question that may arise is whether the gain can be explained by the fact that we have three receivers. Our results, not shown due to space constraints, show that the gain from averaging is not as much as the gain from using the Riemannian tools demonstrating that the proposed algorithms offer an additional advantage beyond the advantage of using multiple receivers.

\begin{figure}[!h]
        \centering
        \begin{subfigure}[b]{0.35\textwidth}
            \centering
        \includegraphics[width=\textwidth]{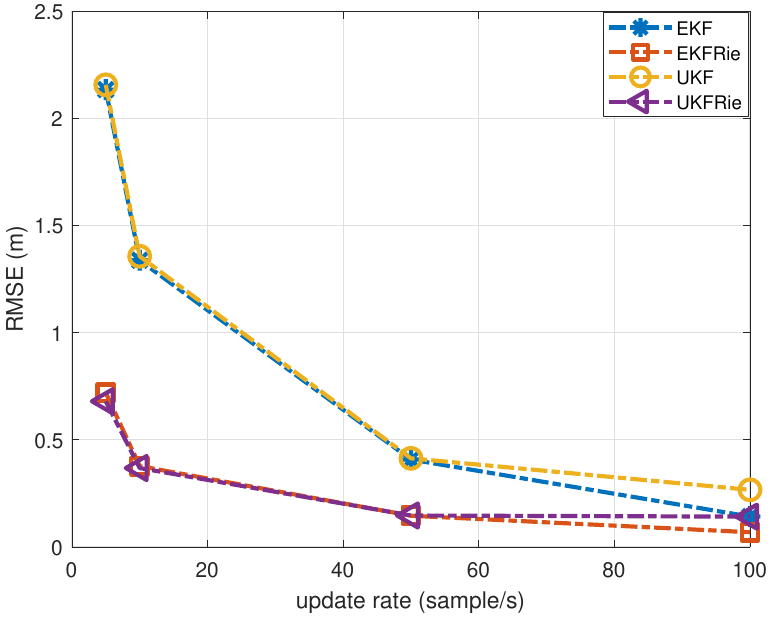}
            \caption[]%
            {{\small Position}}    
        \end{subfigure}
        \hfill
        \begin{subfigure}[b]{0.35\textwidth}
            \centering 
        \includegraphics[width=\textwidth]{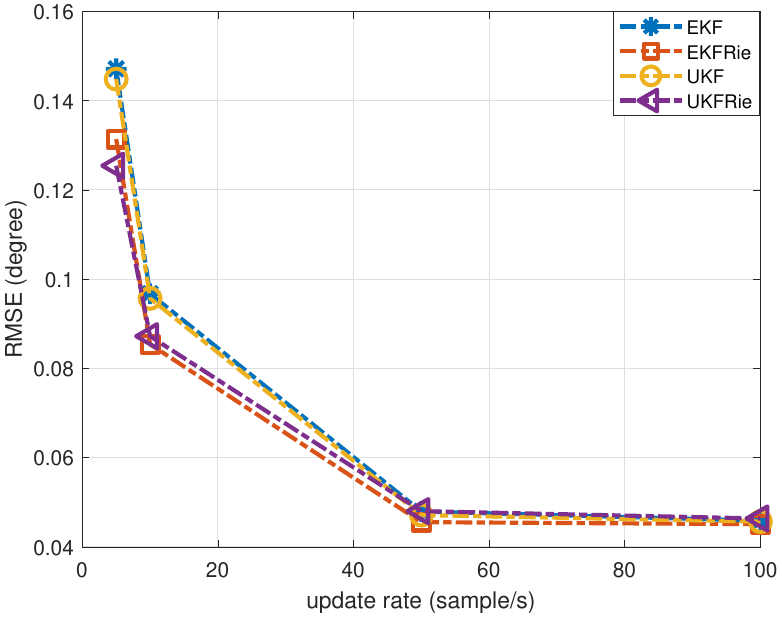}
            \caption[]%
            {{\small Orientation}}   
        \end{subfigure}
        \hfill
        \caption[ ]
        {{\small RMSE of U-path vs IMU's updating rate for $\sigma_r^{2}=1\times 10^{-1}$ }}   \label{fig:RMSEvsUSupdatingRate_path.1_1e-1}
    \end{figure}
\begin{figure}[!ht]
        \centering
        \begin{subfigure}[b]{0.35\textwidth}
            \centering
            \includegraphics[width=\textwidth]{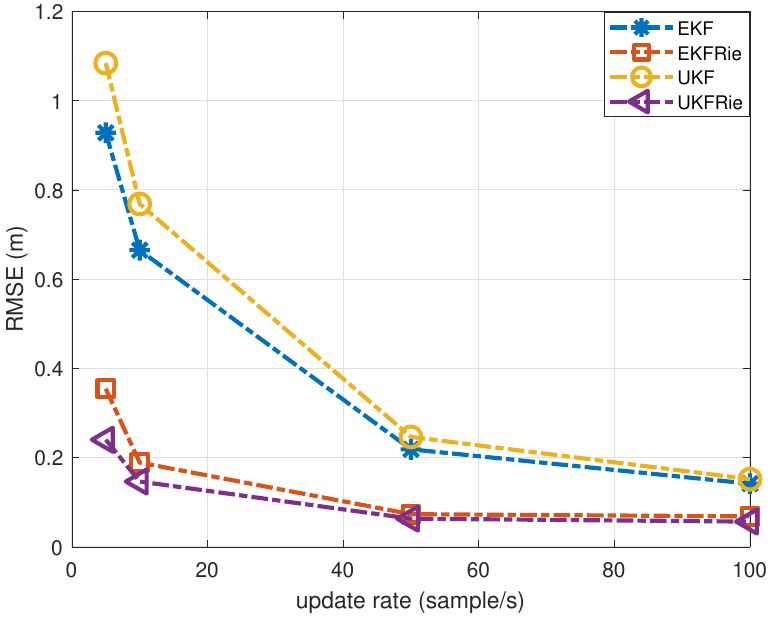}
            \caption[]%
            {{\small Position}}    
        \end{subfigure}
        \hfill
        \begin{subfigure}[b]{0.35\textwidth}  
            \centering 
            \includegraphics[width=\textwidth]{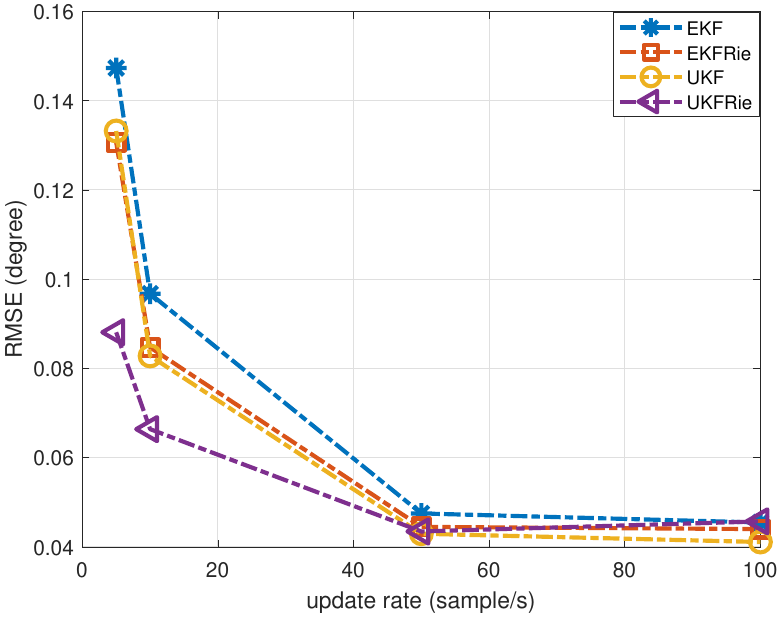}
            \caption[]%
            {{\small Orientation}}   
        \end{subfigure}
        \hfill
        \caption[ ]
        {{\small RMSE of U-path vs IMU's updating rate for $\sigma_r^{2}=1\times 10^{-3}$ }}    
        \label{fig:RMSEvsUSupdatingRate_path.1_1e-3}
    \end{figure}
\begin{figure}[!ht]
        \centering
        \begin{subfigure}[b]{0.35\textwidth}
            \centering
            \includegraphics[width=\textwidth]{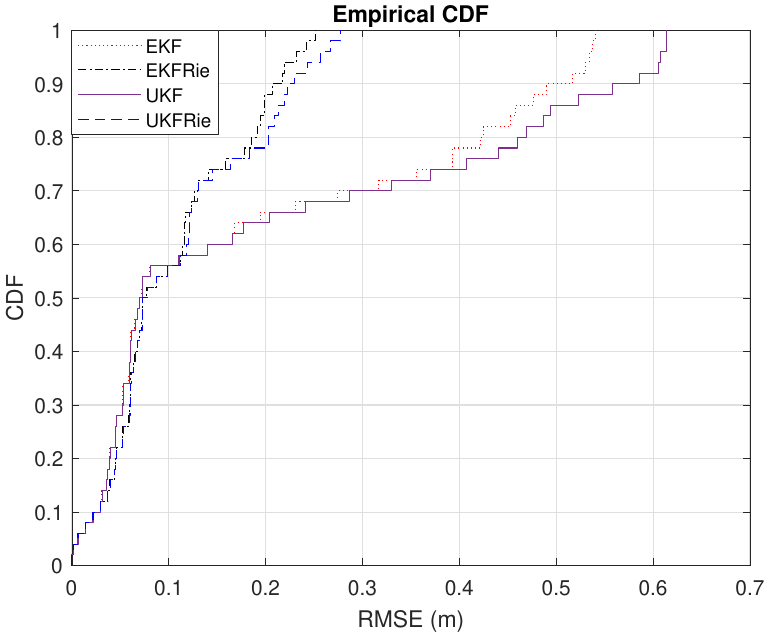}
            \caption[]%
            {{\small $\sigma_r^{2}=1\times10^{-3}$ }}    
        \end{subfigure}
        \hfill
        \begin{subfigure}[b]{0.35\textwidth}  
            \centering 
            \includegraphics[width=\textwidth]{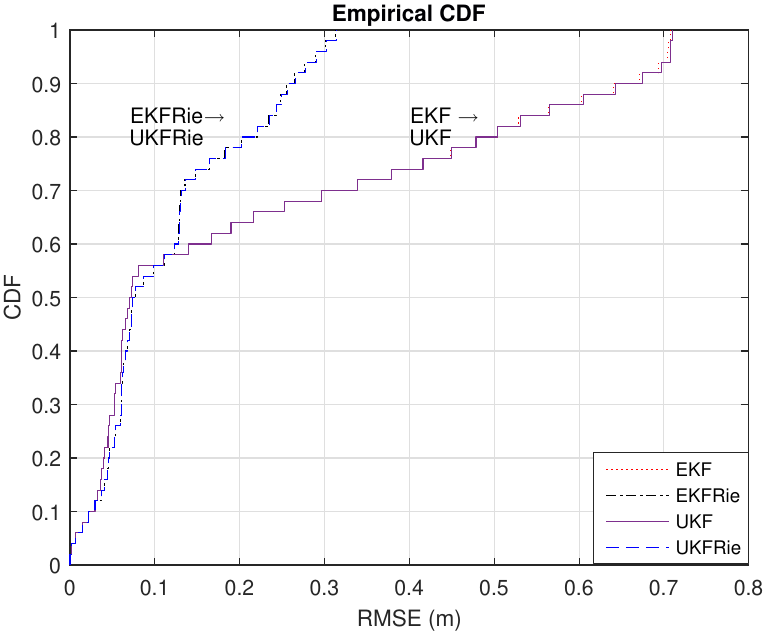}
            \caption[]%
            {{\small $\sigma_r^{2}=1 \times 10^{-1}$}}   
        \end{subfigure}
        \hfill
        \caption[ ]
        {{\small CDF of the RMSE for Bridge-path for different values of  $\sigma_r^{2}$}}
        \label{fig:CDFpath.3}
    \end{figure}

The proposed algorithms show marginal improvements in orientation estimation compared to conventional Kalman filters, as illustrated in Fig. \ref{fig:RMSEvssigma_r_theta}. Fig. \ref{fig:RMSEvsUSupdatingRate_path.1_1e-1} and \ref{fig:RMSEvsUSupdatingRate_path.1_1e-3} show that the increase in IMU rate improves the average accuracy till a certain point at which further increases in IMU rate show diminishing returns. 
Analyzing the cumulative distribution of the RMSE provides a comprehensive understanding of the error distribution and reduces the influence of outliers. Our results show that the proposed algorithms outperform the EKF and UKF across all test cases in terms of position. The CDF of the RMSE of position estimation for Bridge-path at two different of $\sigma_r^{2}$ is shown in Fig. \ref{fig:CDFpath.3}. The CDFs of RMSE for EKFRie and UKFRie at $\sigma_r^{2} = 1\times 10^{-3}$  indicate that $90\%$ of the RMSE lie below $0.22$m and $0.27$m, respectively. In contrast, the corresponding values for EKF and UKF lie below $0.52$m and $0.59$m, respectively.

\begin{figure}[!h]
    \centering
    \includegraphics[width=0.75\linewidth]{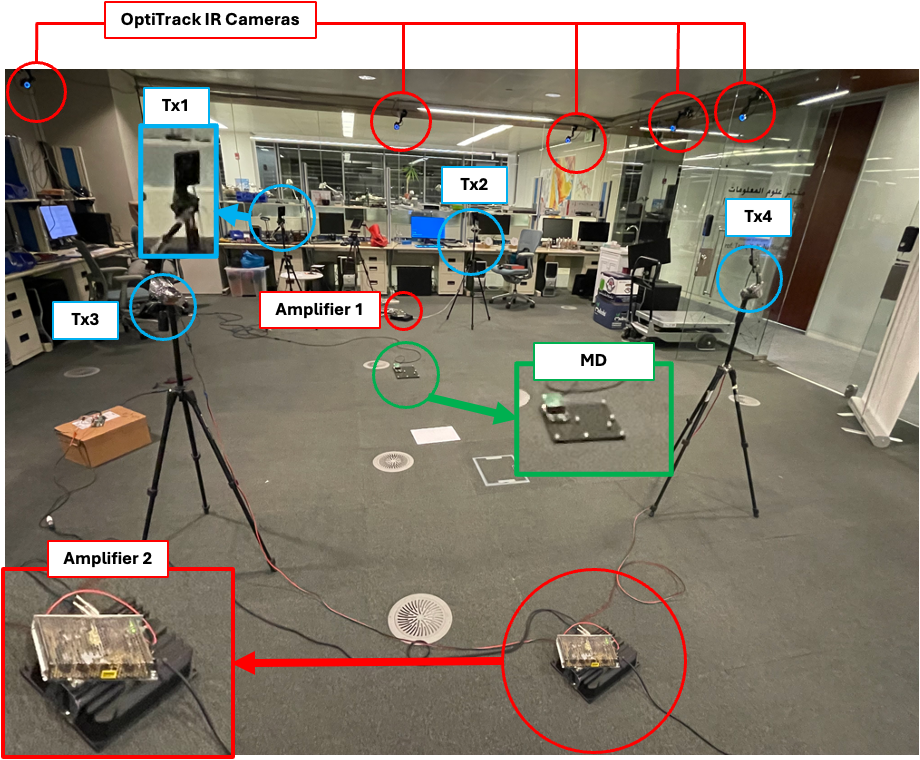}
    \caption{Setup of the experiment \cite{alsharif2024kalman}}
    \label{fig:setup}
\end{figure}
\begin{figure}
    \centering
    \includegraphics[width=0.75\linewidth]{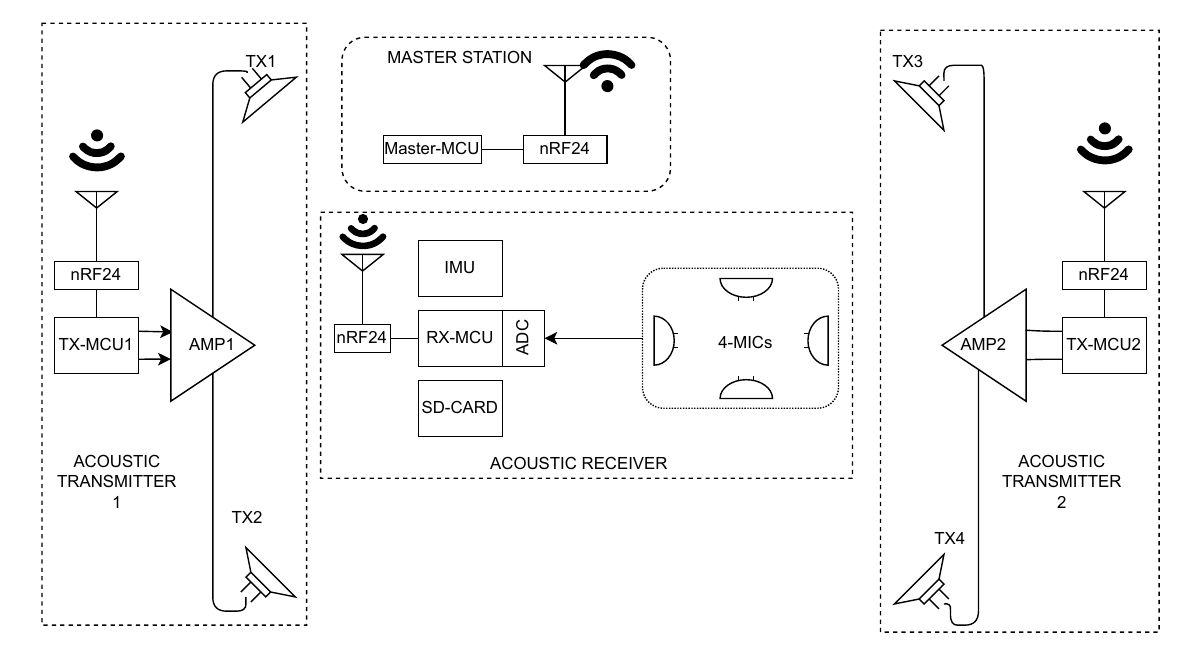}
    \caption{Schematic representation of the experiment setup \cite{alsharif2024kalman}}
    \label{fig:schematic_setup}
\end{figure}
\subsection{Experimental Setup}
In this work we use the same setup as in \cite{alsharif2024kalman}. The experimental setup (shown in Fig. \ref{fig:setup}) consists of four main parts: (a) the optitrack system to provide the ground truth, (b) the mobile device which wishes to localize equipped with IMU  at a data rate of 100 Hz (c) the acoustic transmitters that send an acoustic signal to the mobile device at a data rate of 5 Hz and (d) the master unit that synchronizes the acoustic transmitters and the mobile device. A schematic representation of the setup is shown in Fig \ref{fig:schematic_setup}.

Each of the acoustic transmitters and the mobile device is tagged with optitrack retro-reflective balls that allow the optitrack system to accurately (to about $1$mm) determine its position. The master unit is equipped with a wireless device (NRF24L01) which it uses to send the synchronization signal to the acoustic transmitters and acoustic receivers. Each of the transmitters and mobile device is equipped with the same device (NRF24L01) in receive mode to get the synchronization signals.

The mobile device has 4 microphones in a rhombus shape with sides of length 36.7mm. This provides 4 sets of different triangle configurations that can be used (two equilateral and two isosceles). The IMU, due to engineering constraints, is not aligned to the centroid of any of these four triangles. Thus the readings of the IMU must be transformed to the centroid as explained in the algorithm section. For a further detailed description see \cite{alsharif2024kalman}. 

The considered noise densities of the gyroscope and accelerometer were $0.01 degree/s/\sqrt{Hz}$ and $60 \mu g/\sqrt{Hz}$, respectively, as described in the IMU data sheet \cite{MTi300SpecificationDatasheet}. The variances of the gyroscope and accelerometer were calculated based on the given noise density, as described in the simulation part, and they were $5 \times 10^{-3} degree/s$ and $1.7 \times 10^{-3} m/s^2$. The variance of the acoustic system was determined by comparing its reading with the true reading ($5.5 \times 10^{-3} m^2$).

\subsection{Experimental Results and Discussion} 

In this section, we compare the performance of the proposed algorithms against those proposed by Al-Sharief \textit{et al}. \cite{alsharif2024kalman}. Both approaches utilize Riemannian optimization with EKF and UKF, but with different manifolds and application methods. While Al-Sharief demonstrated that his algorithms outperformed the conventional EKF and UKF, our results show that the algorithms proposed in this work outperform his algorithms.

\begin{figure}[!ht]
        \centering
        \begin{subfigure}[b]{0.35\textwidth}
            \centering
            \includegraphics[width=\textwidth]{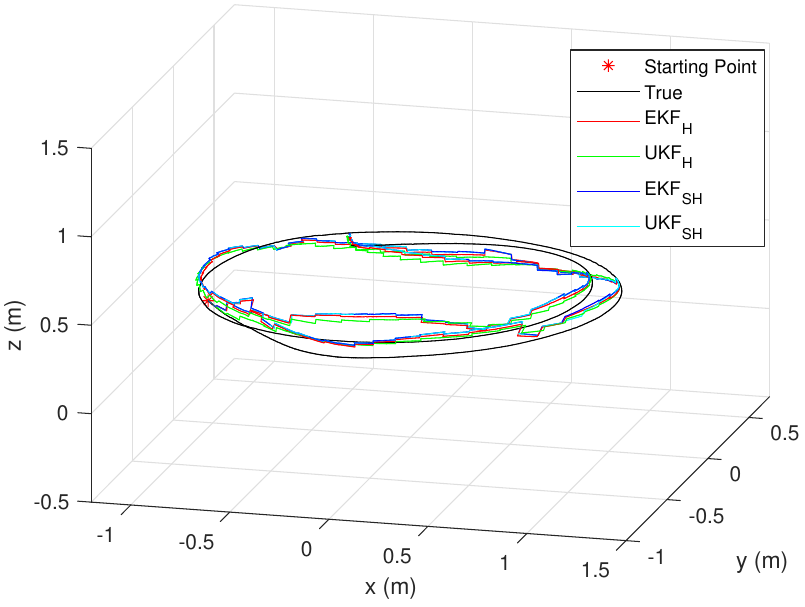}
            \caption[]%
            {{\small 3D }}    
        \end{subfigure}
        \hfill
        \begin{subfigure}[b]{0.35\textwidth}  
            \centering 
            \includegraphics[width=\textwidth]{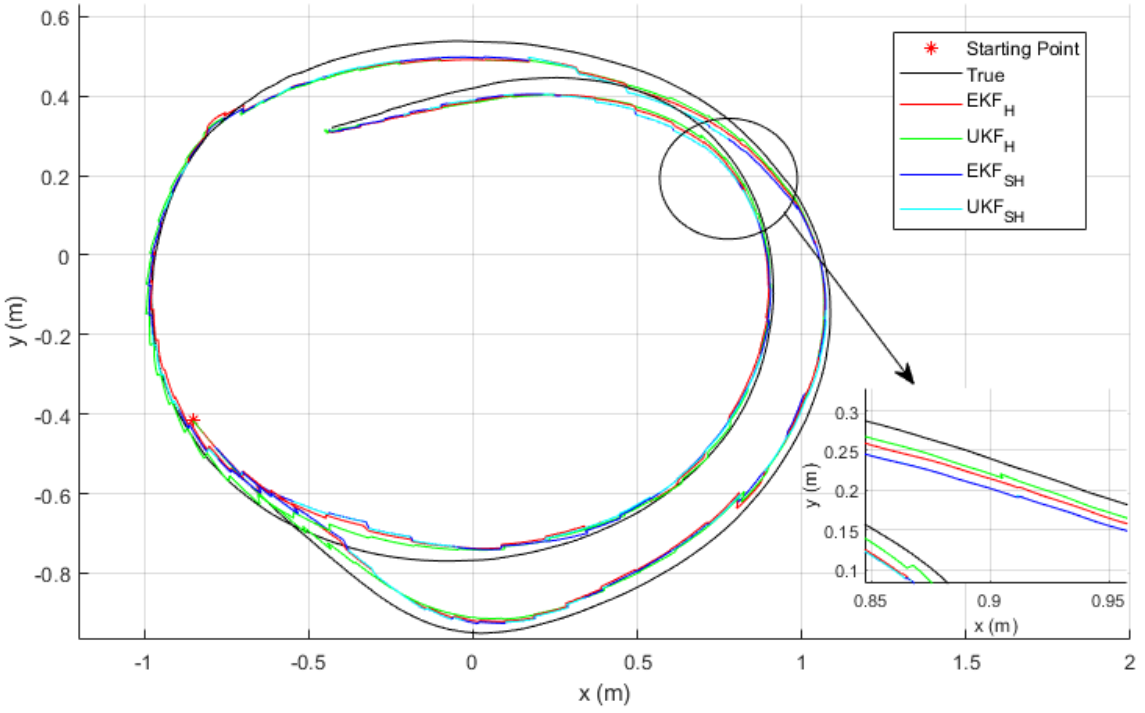}
            \caption[]%
            {{\small Top view}}   
        \end{subfigure}
        \hfill
        \caption[ ]
        {{\small True and estimated path of the experiment}}
        \label{fig:pathOfEperiment}
    \end{figure}

Fig. \ref{fig:pathOfEperiment} illustrates the true and estimated paths generated by our proposed algorithms and those of Al-Sharief (subscript SH). The experiment results demonstrate that the proposed algorithms outperform Al-Sharief's algorithms in terms of both position and orientation, as evidenced by the lower RMSE values in the table. \ref{table:RMSEofNonLinearExp}.

 While the position RMSE of Al-sharief's algorithms was above $7.25 $cm for both EKFRie$_{\text{SH}}$ and UKFRie$_{\text{Sh}}$, the proposed EKFRie and UKFRie satisfied position RMSE equal to $6.73 $cm and $6.16 $cm, respectively. In terms of orientation, Al-Sharief's algorithm satisfied RMSE of $2.1$ degree while the proposed algorithms satisfied RMSE of $1.7$ and  $1.8$ degree for EKFRie and UKFRie, respectively.


\begin{table}[!h]
\centering
\caption{RMSE of the experiment for position (cm) and orientation (degree)}
\label{table:RMSEofNonLinearExp}
\begin{tabular}{|l|l|l|l|l|}
\hline
\multirow{2}{*}{} & \multicolumn{4}{l|}{\hspace{2.2 cm} Algorithms} \\ \cline{2-5} 
  RMSE & EKFRie & UKFRie & EKFRie$_{\text{SH}}$ & UKFRie$_{\text{SH}}$ \\ \hline
 position & \hspace{0.5em} 6.73 &\hspace{0.5em} 6.16 & \hspace{0.5em}7.26 &\hspace{0.5em} 7.27 \\ \hline
  orientation &\hspace{0.6em} 1.7  & \hspace{0.9em}1.8 &\hspace{0.5em} 2.3 &\hspace{0.8em} 2.3 \\ \hline
\end{tabular}
\end{table}

\section{Conclusion}
\label{sec:Concl}

This study explores whether utilizing the SO(3) manifold structure of the rotation matrix to conduct Riemannian optimization can enhance the performance of target tracking in indoor environments. To achieve this, the conventional EKF and UKF were modified by incorporating Riemannian optimization tools, namely the retraction and vector transport, to update the state vector and covariance matrix. Additionally, the impact of adopting the isosceles triangle manifold of the attached transmitters was also examined.

 Our simulation results demonstrated that in static scenarios, incorporating the SO(3) structure did not improve the tracking performance. However, for dynamic scenarios, the proposed EKFRie and UKFRie algorithms outperformed the conventional EKF and UKF in terms of RMSE for position and orientation. While the RMSE for EKF and UKF were $0.36 $m and $0.43 $m for Stair-path, respectively, they were $0.12 $m and $0.10 $m for EKFRie and UKFRie. Additionally, this work showed experimentally the outperforming of the proposed algorithms over the algorithms proposed by \cite{alsharif2024kalman} for both position and orientation. 
 Our results also show that using the isosceles triangle manifold algorithm on the measurements did not improve tracking performance.
 
 One of the future directions is employing the exponential map and parallel transport as alternatives to retraction and vector transport to investigate their potential for performance enhancement.

\section{Acknowledgment}
Hammam would like to thank the Information System Lab (ISL) in KAUST for supporting him in conducting this work. Also, Prof. Muqaibel and Hammam would like to acknowledge King Fahd University of Petroleum and Minerals (KFUPM) for supporting them in this research.


\appendix
\label{sec:Derivation}
This appendix shows the derivation of the formula to transform the acceleration and its variance from the IMU's position in BCS ($\boldsymbol{p}^{b}_{imu}$) into the BCS's centroid ($\boldsymbol{p}^{b}_c$) where $\boldsymbol{z}$ refers to the displacement between $\boldsymbol{p}^{b}_{imu}$ and $\boldsymbol{p}^{b}_c$ and $\mathbf{R}$ refers to the rotation matrix. While position, velocity, and acceleration are time functions, the time symbol has been ignored for convenience.

Let us begin with the following:

\begin{align}
\hspace{-3 cm}\boldsymbol{p}^{b}_{imu} = \boldsymbol{p}^{b}_c  + \mathbf{R} \boldsymbol{z}
\end{align}
so,
\begin{align}
\boldsymbol{a}^{b}_{imu} &= \frac{d^{2}\boldsymbol{p}^{b}_{imu}}{dt^2} \\
    &= \frac{d}{dt} \left[ \frac{d\boldsymbol{p}_{c}^{b}}{dt} + \frac{d}{dt}\mathbf{R}\boldsymbol{z} \right]\\
    &= \frac{d}{dt} \left[ \boldsymbol{v}_{c}^{b} + \frac{d}{dt}\mathbf{R}\boldsymbol{z} \right]\\
    &= \frac{d}{dt} \left[ \boldsymbol{v}_{c}^{b} + \mathbf{R}\boldsymbol{\Omega} \boldsymbol{z} \right]\\
    &= \boldsymbol{a}^{b}_c + \frac{d}{dt} \left[ \mathbf{R}\boldsymbol{\Omega} \boldsymbol{z} \right]\\
    &= \boldsymbol{a}^{b}_c + \left( \boldsymbol{\Omega}\frac{d\mathbf{R}}{dt} + \mathbf{R}\frac{d\boldsymbol{\Omega}}{dt} \right)\boldsymbol{z} \\
    &= \boldsymbol{a}^{b}_c + (\mathbf{R}\boldsymbol{\Omega}^2 + \mathbf{R}\boldsymbol{\alpha})\boldsymbol{z}
    \label{eq.12}
\end{align}
where $\boldsymbol{\Omega}$ defined in \eqref{eq.omeg} and $\boldsymbol{\alpha} = (\boldsymbol{\boldsymbol{\Omega}_f} - \boldsymbol{\boldsymbol{\Omega}_i})/T = \boldsymbol{\Bar{\boldsymbol{\Omega}}}/T$.
We can rearrange \eqref{eq.12} to find $\boldsymbol{a}_{c}^{b}$ such that:
\begin{equation}
\label{eq:a_imu2a_c}
    \boldsymbol{a}^{b}_c = \boldsymbol{a}^{b}_{imu} - (\mathbf{R}\boldsymbol{\boldsymbol{\Omega}}^2 + \mathbf{R}\boldsymbol{\alpha})\boldsymbol{z} 
\end{equation}

Now we will find the variance of $\boldsymbol{a}^{b}_c$ as follows:
\begin{equation}
\label{eq:VAR_aimu2Var_ac}
    Var(\boldsymbol{a}^{b}_c) = Var(\boldsymbol{a}^{b}_{imu}) + Var(\mathbf{R}\boldsymbol{\Omega}^2\boldsymbol{z}) + Var(\mathbf{R}\boldsymbol{\Bar{\boldsymbol{\Omega}}}\boldsymbol{z}/T)
\end{equation}
$Var(\boldsymbol{a}^{b}_{imu})$ is given from the IMU data sheet. The product of $\mathbf{R}\boldsymbol{\Omega}^2\boldsymbol{z}$ and $\mathbf{R}\boldsymbol{\Bar{\boldsymbol{\Omega}}}\boldsymbol{z}/T$ are vectors belong to $\mathbb{R}^{3}$. The $i^{th}$ element of $\mathbf{R}\boldsymbol{\Bar{\boldsymbol{\Omega}}}\boldsymbol{z}/T$ is equal to $\frac{1}{T}\sum_{k=1}^{3}z_{k}\sum_{j=1}^{3}(r_{ij}\bar{\omega}_{jk})$ where $r_{ij}$ and $\bar{\omega}_{ij}$ refer to the element in $i^{th}$ row and $j^{th}$ column in $\mathbf{R}$ and $\boldsymbol{\bar{\Omega}}$, and $z_k$ refers to the $k^{th}$ element in $\boldsymbol{z}$.

We can find the variance of the $i^{th}$ element of $\mathbf{R}\boldsymbol{\Bar{\boldsymbol{\Omega}}}\boldsymbol{z}/T$ as:
    \begin{equation}
    \begin{split}
   Var\left( i^{th} \text{ element of }\frac{\mathbf{R}\boldsymbol{\Bar{\boldsymbol{\Omega}}}\boldsymbol{z}}{T}\right) &=  \\ \frac{1}{T^{2}}\sum_{k=1}^{3}z^{2}_{k}\sum_{j=1}^{3}Var&(r_{ij}\bar{\omega}_{jk})
\label{eqn2}
    \end{split}
\end{equation}

 By assuming independence between $r_{ij}$ and $\bar{\omega}_{jk}$ for simplicity, we find $Var(r_{ij}\Bar{\omega}_{jk})$ as follows, where $E$ refers to the expectation:
\begin{align*}
    Var&(r_{ij}\Bar{\omega}_{jk}) =  Var(r_{ij})Var(\Bar{\omega}_{jk}) + \\ &Var(r_{ij})E[\Bar{\omega}_{jk}]^2 + Var(\Bar{\omega}_{jk})E[r_{ij}]^2 \numberthis 
    \label{eqn1}
\end{align*}

The estimated $r_{ij}$ and $\Bar{\omega}_{jk}$ from the algorithms are used as the expectations. The variance of $r_{ij}$ is extracted from the covariance matrix of the Kalman filter while the variance of $(\Bar{\omega}_{jk})$ is equal to twice the variance of the angular velocity which is given in the IMU's data sheet.

Regarding the other term ($\mathbf{R}\boldsymbol{\Omega}^2\boldsymbol{z}$), let us find the mean and variance of $\boldsymbol{\Omega}^2$ first, where:
\begin{equation}
    \boldsymbol{\Omega}^2 = \begin{bmatrix}
    -\omega^{2}_z - \omega^{2}_y & \omega_x\omega_y & \omega_x\omega_z \\
    \omega_x\omega_y & -\omega^{2}_z - \omega^{2}_x & \omega_z\omega_y \\
    \omega_x\omega_z & \omega_y\omega_z & -\omega^{2}_y - \omega^{2}_x
\end{bmatrix} 
\end{equation}
By assuming independence, we can find $Var(\omega_i\omega_j)$ by the same formula used in \eqref{eqn1}, and $E[\omega_i\omega_j] = E[\omega_i]E[\omega_j]$. To find mean and variance of $\omega_i^{2}$, the author uses the statistics for the square of variables such that:
\begin{align}
    E(\omega_{i}^2)=& Var(\omega_{i})+E[\omega_{i}]^2\\
    Var(\omega_{i}^2) =& 2Var(\omega_{i})^2 + 4E[\omega_{i}]Var(\omega_{i})
\end{align}
Finally, we can find the variance of $i^{th}$ element of $\mathbf{R}\boldsymbol{\Omega}^2\boldsymbol{z}$ by using \eqref{eqn2} with replacing $\Bar{\boldsymbol{\Omega}}$ with $\boldsymbol{\Omega}^2$ and deleting the time period ($T$). 


\bibliographystyle{unsrt}
\bibliography{main}



\ifCLASSOPTIONcaptionsoff
  \newpage
\fi

\end{document}